\documentclass[11pt,a4paper,onecolumn]{article}
\usepackage[T1]{fontenc}
\usepackage[utf8]{inputenc}
\usepackage{authblk}
\usepackage{graphicx}
\usepackage{subfigure} 
\makeatletter
\renewcommand{\@thesubfigure}{\normalsize(\textbf{\alph{subfigure}})}
\makeatother
\usepackage{siunitx}
\newcommand{\pg}{\ensuremath{(\mathrm{p},\gamma)}}
\newcommand{\pn}{\ensuremath{(\mathrm{p},\mathrm{n})}}
\newcommand{\pa}{\ensuremath{(\mathrm{p},\alpha)}}
\newcommand{\alal}{\ensuremath{(\alpha,\alpha)}}
\newcommand{\an}{\ensuremath{(\alpha,\mathrm{n})}}
\newcommand{\ag}{\ensuremath{(\alpha,\gamma)}}

\newcommand{\ion}[3]{\ensuremath{{}^{#1}\mathrm{#2}^{#3}}}
\newcommand{\nuc}[2]{\ion{#1}{#2}{}}

\newcommand{\react}[5]{\ensuremath{\nuc{#1}{#2}#3\nuc{#4}{#5}}}

\newcommand{\Ep}{\ensuremath{E_\mathrm{p}}}
\newcommand{\Ea}{\ensuremath{E_\alpha}}
\newcommand{\Eg}{\ensuremath{E_\gamma}}
\newcommand{\Ecm}{\ensuremath{E_\mathrm{cm}}}

\newcommand{\Ex}{\ensuremath{E_\mathrm{x}}} 

\newcommand{\Jpi}{\ensuremath{J^\pi}}

\title{Underground Measurements of Nuclear Reaction Cross-Sections Relevant to AGB Stars}
\date{}

\author{
Chemseddine Ananna~$^{1,2}$,
Francesco Barile~$^{3,4}$,
Axel Boeltzig~$^{1,2\dagger}$,
Carlo Giulio Bruno~$^{5}$,
Francesca Cavanna~$^{6,}$,
Giovanni Francesco Ciani~$^{3,4}$,
Alessandro Compagnucci~$^{7}$,
Laszlo Csedreki~$^{8}$,
Rosanna Depalo~$^{9,10,}$,
Federico~Ferraro~$^{9,10}$,
Eliana Masha~$^{9,10}$,
Denise Piatti~$^{11,12}$,
David Rapagnani~$^{1,2}$ and 
Jakub Skowronski~$^{11,12}$
}

\affil{%
$^{1}$ Universit\`a di Napoli Federico II, Dipartimento di Fisica E. Pancini, Strada Comunale Cintia, 80126 Napoli, Italy\\ %MDPI: Affs. 1, 2, 7, 8, Please add the department/school/faculty/campus.
$^{2}$ INFN, Sezione di Napoli, Strada Comunale Cintia, 80126 Napoli, Italy;\\

$^{3}$ Universit\`a degli Studi di Bari, Dipartimento Interateneo di Fisica M. Merlin, Via Edoardo Orabona 4, 70125 Bari, Italy \\
$^{4}$ INFN, Sezione di Bari, Via Edoardo Orabona 4, 70125 Bari, Italy;\\
$^{5}$ SUPA, School of Physics and Astronomy, University of Edinburgh, Edinburgh EH9 3FD, UK\\
$^{6}$ INFN, Sezione di Torino, Via Pietro Giuria 1, 10125 Torino, Italy \\
$^{7}$ Gran Sasso Science Institute, INFN, Viale Francesco Crispi 7, 67100 L’Aquila, Italy\\
$^{8}$ Institute for Nuclear Research (ATOMKI), P.O. Box 51, HU-4001 Debrecen, Hungary\\
$^{9}$ Universit\`a degli Studi di Milano, Dipartimento di Fisica, Via Giovanni Celoria 16, 20133 Milano, Italy \\
$^{10}$ INFN, Sezione di Milano, Via Giovanni Celoria 16, 20133 Milano, Italy;
$^{11}$ Universit\`a degli Studi di Padova, Dipartimento di Fisica e Astronomia G. Galilei, Via Francesco Marzolo 8, 35131 Padova, Italy\\
$^{12}$ INFN, Sezione di Padova, Via Francesco Marzolo 8, 35131 Padova, Italy;\\

$^{\dagger}$ Current address: Helmholtz-Zentrum Dresden-Rossendorf, Bautzner Landstraße 400, 01328 Dresden, Germany;\\
}
\begin{document}
%%%%%%%%%%%%%%%%%%%%%%%%%%%%%%%%%%%%%%%%%%
\maketitle

\abstract{Nuclear reaction cross sections are essential ingredients to predict the evolution of AGB stars and understand their impact on the chemical evolution of our Galaxy. Unfortunately, the cross sections of the reactions involved are often very small and challenging to measure in laboratories on Earth. In this context, major steps forward were made with the advent of underground nuclear astrophysics, pioneered by the Laboratory for Underground Nuclear Astrophysics (LUNA). The present paper reviews the contribution of LUNA to our understanding of the evolution of AGB stars and related nucleosynthesis.}

\section{Introduction} % Rosanna
The study of Asymptotic Giant Branch stars (AGB) and related nucleosynthesis is crucial to understand the chemical evolution of our Galaxy~\cite{Busso99-AnnRev}. Measuring nuclear cross sections directly at the energies of interest for AGB stars, and for stars in general, is a challenge that sparked big experimental progresses. At typical stellar temperatures, the kinetic energy of interacting nuclei is much lower than the Coulomb repulsion energy. Therefore nuclear reactions among charged particles can only occur via quantum-mechanical tunneling, which has a very small probability. As a result, the nuclear cross sections involved are usually small and difficult
to measure. Direct cross section measurements typically require accelerators providing high-intensity beams, ultra-pure stable targets, and the highest possible detection efficiency. Moreover, when approaching the astrophysically relevant energy range, the counting rate from a nuclear fusion reaction signal may become much smaller than the environmental background produced by cosmic radiation and naturally-occurring radioactive elements.

A breakthrough in experimental nuclear astrophysics was introduced by the construction of the Laboratory for Underground Nuclear Astrophysics (LUNA) in Gran
Sasso, Italy. Over the last 30 years, the LUNA collaboration has provided cross section data directly at the relevant energies for AGB nucleosynthesis for a number of reactions. 

The next sections provide a description of the experimental equipment available at LUNA, as well as a summary of the latest results. We will focus in particular on the implications of the new results on AGB stellar evolution and nucleosynthesis, although some of the nuclear reactions may impact also other astrophysical scenario (such as core hydrogen burning or classical novae explosions). 

%%%%%%%%%%%%%%%%%%%%%%%%%%%%%%%%%%%%%%%%%%
\section{The Laboratory for Underground Nuclear Astrophysics}

\begin{figure}[h]
\begin{center}
    \includegraphics[width=10.cm]{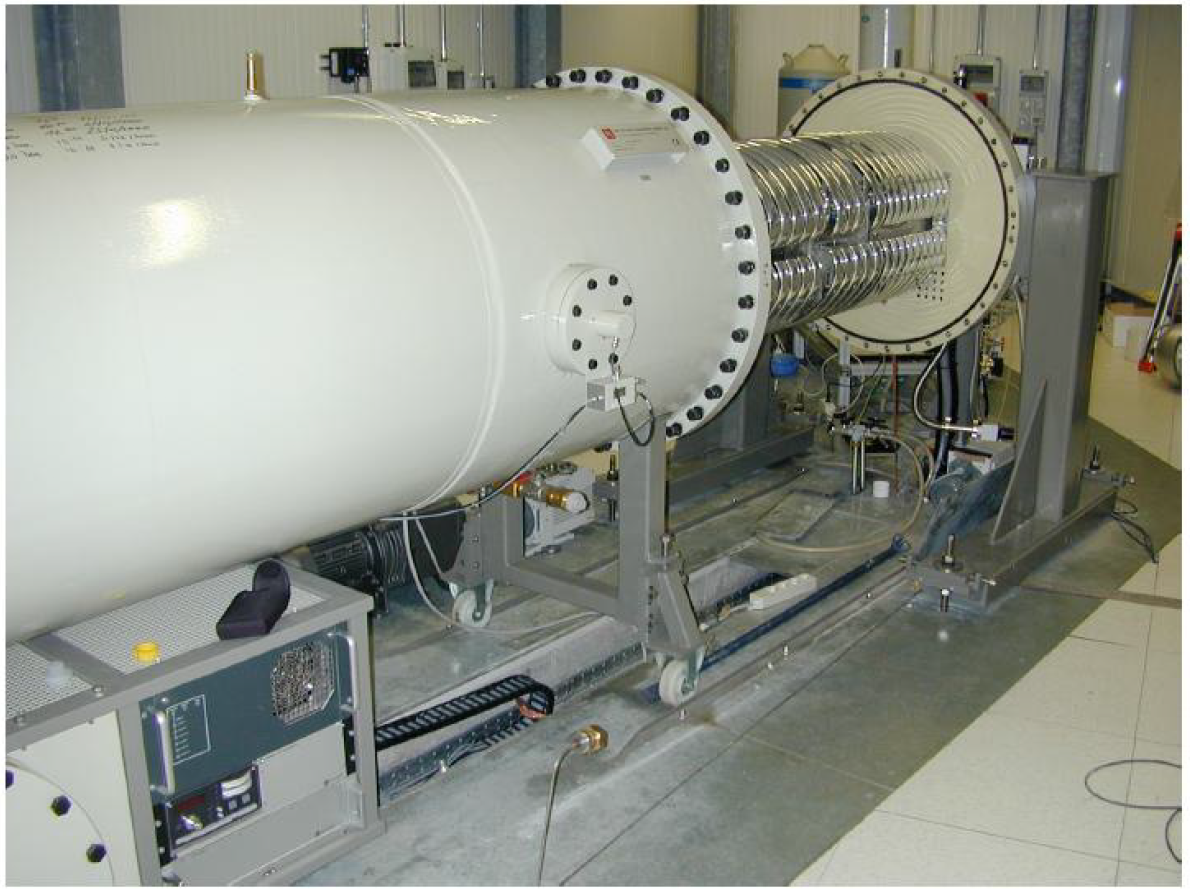}
    
    \caption{Open tank of the LUNA \SI{400}{kV} accelerator. Inside the tank: the metallic rings keeping the ion source area free from electric fields, are visible.}
    \label{fig:tank}
\end{center}
\end{figure}

LUNA is located at the INFN Gran Sasso National Laboratory (LNGS), Italy, where the \SI{1400}{m} of rocks covering the laboratory guarantee a reduction of six (three) orders of magnitude in cosmic muon (neutron) flux.  The first accelerator installed underground at LNGS was a compact \SI{50}{kV} machine~\cite{Greife1994}, now decomissioned, used to perform the first measurements in 1994. In the year 2000, a commercial high-current \SI{400}{kV} accelerator from High Voltage Engineering Europe, was installed and is actively used for nuclear astrophysics experiments up to this day. Figure~\ref{fig:tank} shows the tank of the LUNA \mbox{\SI{400}{kV} accelerator}.
The electrostatic accelerator is embedded in a steel tank (\SI{1}{m} diameter and \SI{2.8}{m} long), which is filled with a mixture of N$_2$ and CO$_2$ gas at a total pressure of \SI{20}{bar}. The high voltage (HV) is generated by an inline-Cockcroft--Walton power supply (located inside the tank); it is stabilized by a RC-filter at the HV power supply output and by an active feedback loop based on a chain of resistors. 
The accelerator provides very stable beams of protons and alpha-particles in the range {50}--{400}{keV} with a maximum current on-target up to \SI{500}{\micro A}, an energy spread of \SI{0.1}{keV} and a long term stability of \SI{5}{eV} per hour. The beam is provided by a radio frequency ion source from the excitation of a gas that forms an ion plasma with charge e${^+}$ confined by an axial magnetic field. The source is mounted directly on the accelerator tube and the ions are extracted by a voltage applied to an electrode inside the tube itself~\cite{Formicola2003}.
The ions can be directed into one of two beam lines, thereby allowing the installation of two different target setups, used to install a solid target on one, and a gas target on the other beam line. In the first case, the proton beam is guided and focused to the target station using an highly stable analysing magnet, steering magnets and quadrupoles. Before reaching the target, the beam passes through a copper pipe cooled to liquid nitrogen temperature and typically biased to \SI{-300}{V}. The pipe serves both as a cold trap to prevent carbon buildup on the target, and as a secondary electron suppressor for accurate beam current reading. The beam is stopped in the target backing, which is directly cooled. 
Instead, in case of the windowless gas target, an analysing magnet and a steering magnet guide the beam through three pumping stages connected by collimators of increasingly small diameter. Vacuum pumps gradually decrease the pressure from a few mbar inside the target chamber to the level of $10^{-8}\,\mathrm{mbar}$ inside the accelerating tube. When the ion beam passes through the gas target chamber many secondary electrons are released, making the electrical current reading unreliable. For this reason, the beam is stopped on a beam power calorimeter that provides a precise determination of the beam current~\cite{Cavanna2014,Ferraro2018-EPJ,Mossa2020-EPJA}. 
For cross section measurements of radiative capture reactions at LUNA, three high-energy resolution High Purity Germanium detectors (HPGe) and a high gamma detection efficiency BGO (Bi$_4$Ge$_3$O$_{12}$) crystal are currently used. The latter is made of an array of six prismatic crystals, each \SI{28}{cm} long and with a trapezoidal base area of $\SI{52}{cm^2}$. The segments are arranged around a cylindric borehole, so that the target can be placed at the centre of the detector with a radial thickness of at least \SI{7}{cm} of BGO around. To further reduce environmental background, different lead shieldings have been designed for use with either the BGO or the HPGe detector~\cite{Cavanna2014,Boeltzig18-JPG}.

In addition, the collaboration has developed an array of silicon detectors for charged particles~\cite{Bruno2015} and a neutron counter based on \nuc{3}{He} tubes~\cite{Balibrea2018,Csedreki2021}.

The LUNA Collaboration is now entering a new phase, which will offer the possibility to study also nuclear reactions occurring in more advanced phases of stellar evolution, such as helium burning and carbon burning. For this purpose, a new \SI{3.5}{MV} accelerator called LUNA~MV is presently being installed at the north side of Hall~B at LNGS. 
The accelerator room, with a surface of about \SI{300}{m^2}, will have \SI{80}{cm} thick concrete walls and ceiling working as neutron shielding towards the outside of the experimental area. The maximum additional neutron flux due to activities at the accelerator is guaranteed to be below the environmental one~\cite{Cavanna18-IJMPA,Broggini2019,Ferraro2021}.
The machine is an Inline Cockcroft Walton accelerator from High Voltage Engineering Europe, that allows to cover a Terminal Voltage range from \num{0.2} to \SI{3.5}{MV} and to deliver ion beams of \ion{}{H}{+}, \ion{4}{He}{+}, \ion{12}{C}{+} in the energy range from \num{0.35} to \SI{3.5}{MeV}. In the case of the carbon beam, using \ion{12}{C}{++} a maximum beam energy of \SI{7}{MeV} can be reached~\cite{Sen2019}. A key feature to perform experiments on reactions important in astrophysics scenarios is the intensity of the beam delivered to the target. The LUNA~MV accelerator will deliver beam currents as high as \SI{1}{emA} for\ion{}{H}{+}, 500 e$\mu$A for \ion{4}{He}{+}, 150 e$\mu$A for \ion{12}{C}{+} and 100 e$\mu$A for \ion{12}{C}{++}. 

The scientific program of LUNA~MV will start with a high-energy study of the \react{14}{N}{\pg}{15}{O} reaction, and will then proceed with key reactions of helium and carbon burning that determine and shape the evolution of massive stars toward their final state, as well as the synthesis of heavy elements through the slow neutron capture process ($s$-process, see, e.g.,~\cite{B2FH}). 

%%%%%%%%%%%%%%%%%%%%%%%%%%%%%%%%%%%%%%%%%%
\section{Hydrogen Burning}
\subsection{CNO Cycle}

\begin{figure}[h!]
 \begin{center}
    \includegraphics[width=0.9\textwidth]{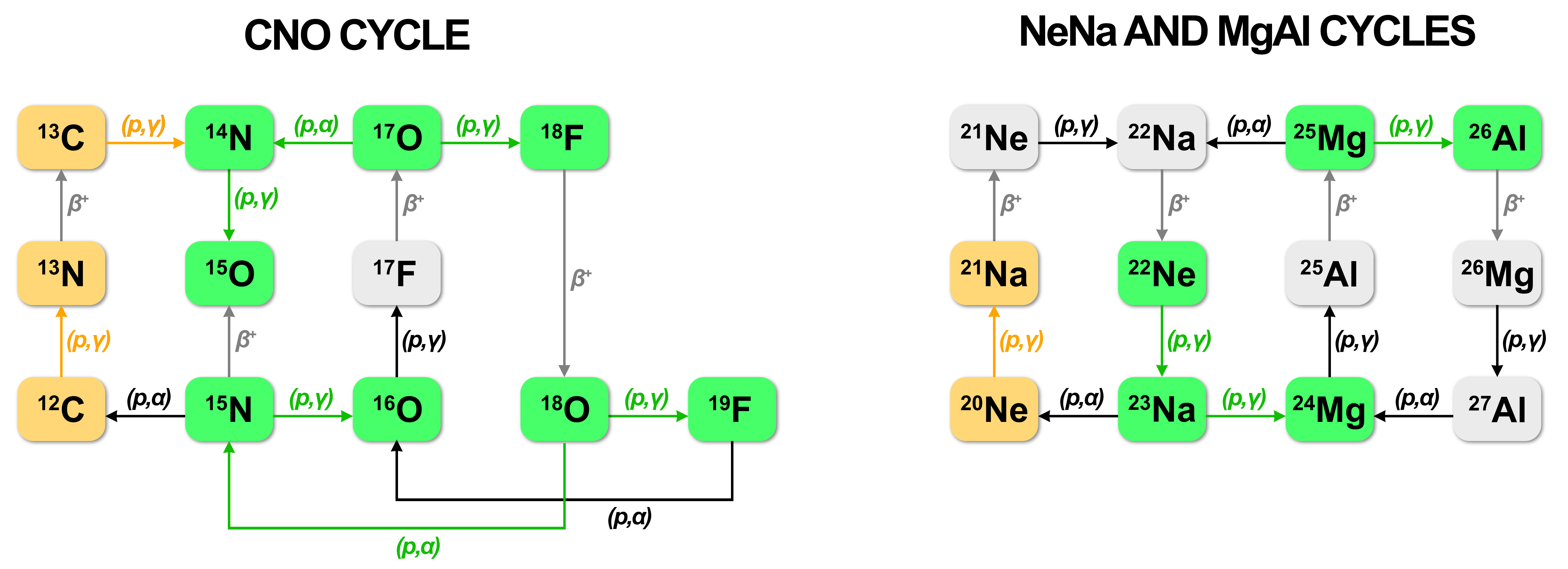}
    \caption{Sketch of the CNO, NeNa and MgAl cycles of hydrogen burning. The reactions highlighted in green have all been studied at LUNA, while the reactions in yellow are presently under investigation.}
    \label{fig:cycles}
\end{center}
\end{figure}

%%General structure for all the sections (max length 1 pag. and 1 figure):
%Specific astrophysical input of each reaction (general notes on AGB stars will be included in a different paper)
%Experimental setup, with focus on innovative aspects (if any). A general description of the solid and gas target beam lines will go to Section 2.
%Results (if any) and/or future prospects
Shell hydrogen burning in AGB stars mainly happens through the CNO cycle. Over the years, the LUNA Collaboration has measured the cross section of most of the CNO cycle reactions directly at the energies of interest for AGB nucleosynthesis~\cite{Broggini2019,Cavanna18-IJMPA} (see \mbox{Figure~\ref{fig:cycles}}). 

In the forthcoming paragraphs we will highlight the most recent results and the experiments in progress.

\subsubsection{\react{17}{O}{\pa}{14}{N}} % Carlo
Pre-solar grains retrieved from meteorites fallen on Earth offer an outstanding probe of the isotopic composition of the stellar sites they condensed in. High-precision measurement of their isotopic ratios is possible in Earth-based laboratories, however matching an isotopic fingerprint to the correct stellar site is often a challenge. The opposite is also true. For example, intermediate-mass (4--8 M$_{\rm Sun}$) AGB stars are expected to contribute significantly to the dust inventory of our Solar System---yet for a long time no grains were identified that could be firmly matched with the predicted isotopic ratios from these stars. In particular, the expected \nuc{17}{O}/\nuc{16}{O} content did not match any candidate grains. This ratio is strongly affected by the \react{17}{O}{\pa}{14}{N} reaction, at the branching point between CNO-II and CNO-III. At temperatures of interest for CNO cycle operation (\SIrange{30}{100}{MK}), this reaction is completely dominated by an $\Ecm~=~\SI{65}{keV}$ resonance, for which direct measurements~\mbox{\cite{Berheide92,Niemeyer96,Blackmon95,Hannam99}} were not in good agreement with each other. Indirect measurements using the Trojan Horse Method were also performed by \cite{Sergi15}.  %MDPI: We changed hyphen into em dash, please confirm -> Yes, Is it ok;

\begin{figure}[h!]
 \begin{center}
    \includegraphics[width=0.4\textwidth]{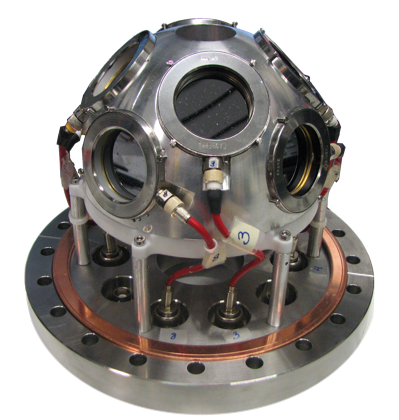}
    \caption{The reaction chamber used in this study (adapted {from}~\cite{Bruno2016}). The ion beam enters from the hole in the top, and hits the solid Ta$_2$O$_5$ target located in the centre of the dome (not shown). Alpha particles are detected at backward angles by an array of eight \mbox{silicon detectors}.} %MDPI: Please confirm if copyright permission is needed for this figure, please check all figures
    \label{fig:pa_chamber}
\end{center}
\end{figure}

The measurement performed underground at LUNA was the first to exploit the background reduction for charged particles at low energy ($\le$$\SI{1}{MeV}$) with an in-beam experiment. Background reduction depends on the energy of the detected charged particles and is typically at least a factor 10~\cite{Bruno2015}. Note this background reduction is due to the reduced flux of gamma rays interacting with the charged-particle detectors via Compton scattering. A proton beam of \SI{150}{\micro A} was accelerated on a solid Ta$_2$O$_5$ target, enriched in \nuc{17}{O}~\cite{Caciolli2012}, and alpha particles were detected by an array of eight silicon detectors at backward angles (\mbox{Figure \ref{fig:pa_chamber}}). Thin aluminised Mylar foils were mounted to shield the detectors from the intense flux of elastically scattered protons. After a lengthy campaign, the strength of the $\Ecm~=~\SI{65}{keV}$ resonance was found to be $\omega\gamma~=~(10.0\pm1.4_{\rm stat}\pm0.7_{\rm syst})$\,neV, almost a factor of two higher than previously estimated. Further details on the analysis are reported in \cite{Bruno2016}. The updated reaction rate was found to be a factor two higher than previously accepted, leading to important consequences in the production of \nuc{17}{O} in low-mass RGB and AGB stars~\cite{Straniero2017}. In particular, using this rate it was possible to come to a very good agreement between predictions from intermediate-mass AGB stars and observations in Group-II silicate grains~\cite{Lugaro2017} under the assumption of \nuc{18}{O} dilution. It is worth noting that more recently an alternative stellar model~\cite{Palmerini2021} proposed these grains come from low-mass AGB stars, assuming deep extra-mixing processes~\cite{Busso2010} but without the need of dilution with solar matter.

\subsubsection{\react{17}{O}{\pg}{18}{F}}
\begin{figure}[h!]
 \begin{center}
    \includegraphics[width=10.cm]{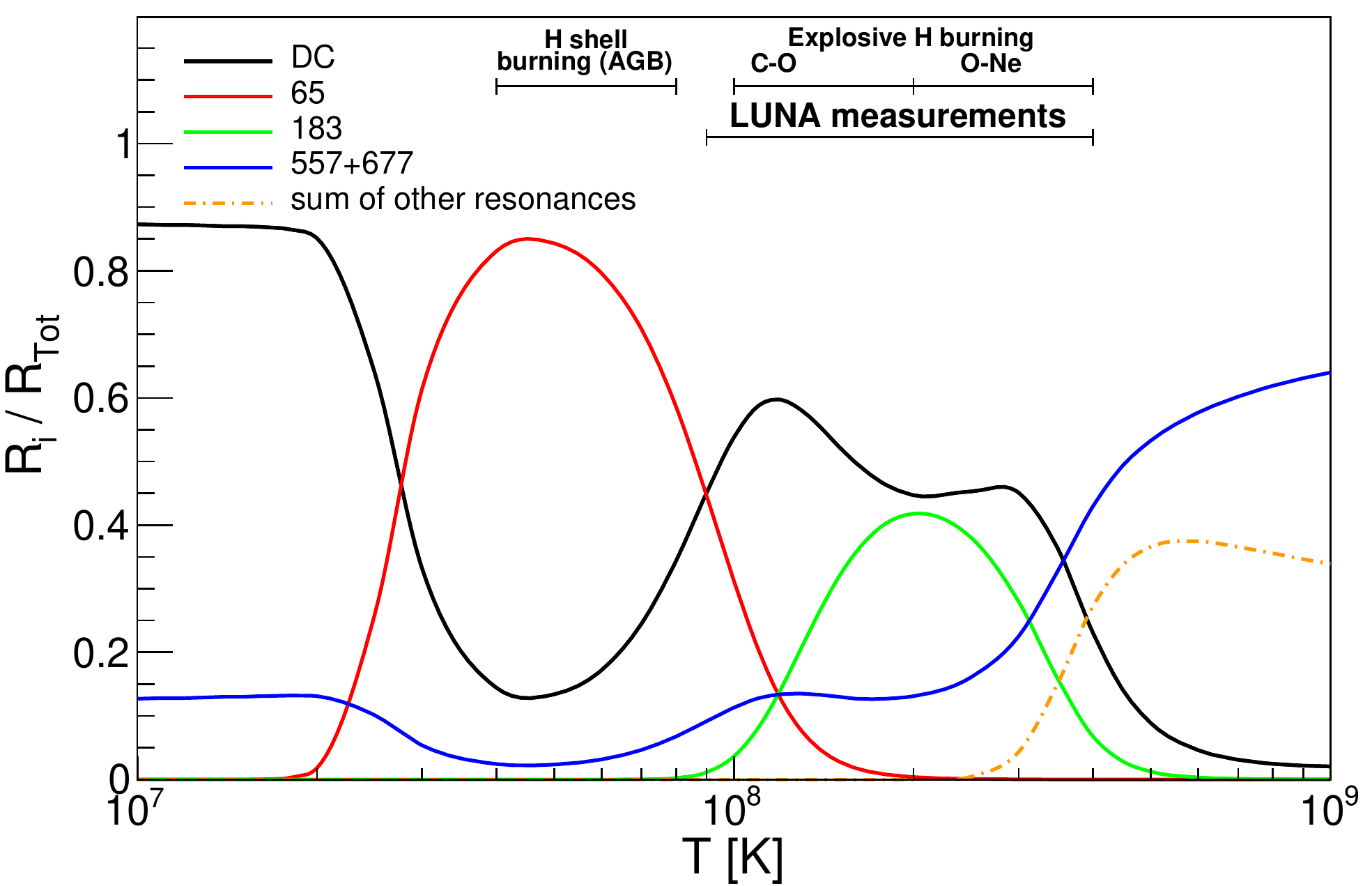}
    \caption{Fractional contribution to \react{17}{O}{\pg}{18}{F} reaction rate.}
    \label{fig1_17Opg}
\end{center}
\end{figure}

{17O(p,gamma)18F} % Carlo, Denise, Giovanni
The \react{17}{O}{\pg}{18}{F} reaction plays a central role in both AGB stars and nova explosions. 
In particular, during a nova runaway the \react{17}{O}{\pg}{18}{F} and \react{17}{O}{\pa}{14}{N} reactions determine the production of the short-lived radioisotope \nuc{18}{F} ($t_{1/2}~=~\SI{110}{min}$) of key interest for nova nucleosynthesis models. During the thermonuclear runaway (\SIrange{100}{400}{MK}), the rate of the \react{17}{O}{\pg}{18}{F} reaction is dominated by the combination of direct capture (DC) contribution and an $\Ecm~=~\SI{183}{keV}$ resonance (Figure~\ref{fig1_17Opg}). Measuring this resonance was the goal of a recent experimental campaign at LUNA~\cite{Scott2012,DiLeva2014}. A \SI{200}{\micro A} proton beam was accelerated onto Ta$_2$O$_5$ solid target enriched in \nuc{17}{O}, and gamma rays were detected with a large-volume HPGe detector in close geometry with the target. This detector was surrounded by a \SI{5}{cm} lead shield to further reduce natural background. After irradiation, targets were taken to the ultra-low counting facility STELLA~\cite{Laubenstein04} at LNGS for activation measurement. The strength of the $\Ecm~=~\SI{183}{keV}$ resonance was measured with unprecedented precision $\omega\gamma$~=~(1.67 $\pm$ 0.12)\,$\mu$eV, and several previously undetected transitions were observed, providing improvements on nova models.

\begin{figure}[h!]
 \begin{center}
    \includegraphics[width=0.32\textwidth]{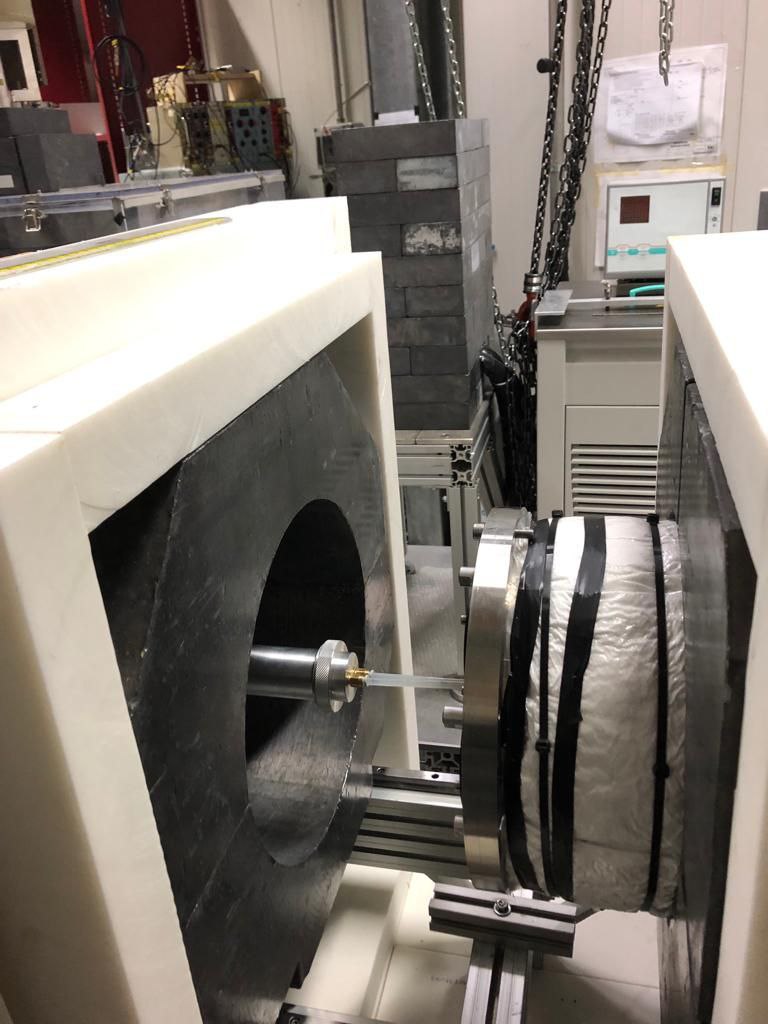}
     \includegraphics[width=0.57\textwidth]{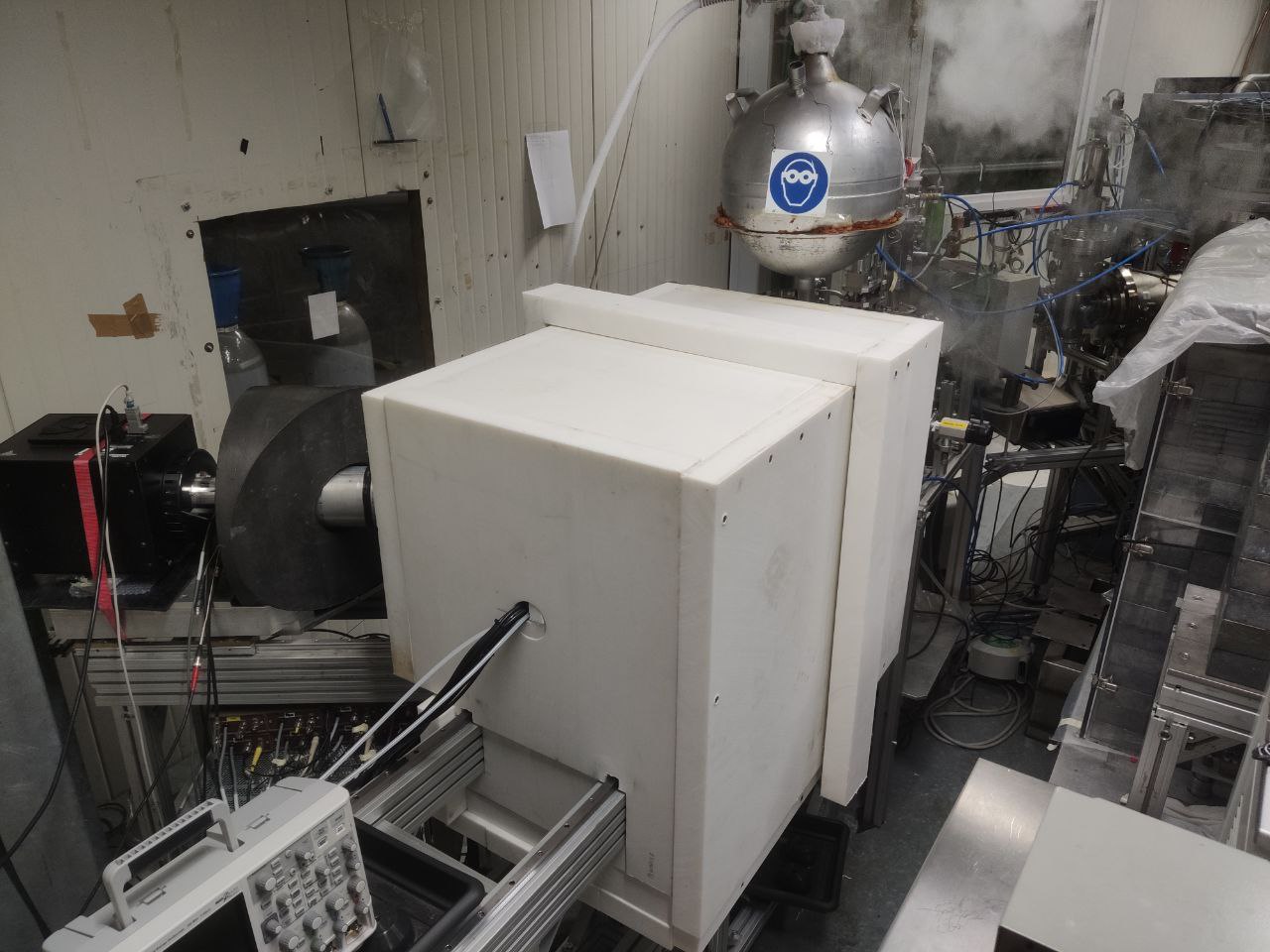}

  \caption{The planned setup for the next LUNA campaign on the \react{17}{O}{\pg}{18}{F} reaction.}
     \label{fig2_17Opg}
\end{center}
\end{figure}

In the temperature range of interest for hydrostatic H burning (\SIrange{40}{80}{MK}) the \react{17}{O}{\pg}{18}{F} reaction rate is dominated by a poorly constrained resonance at $\Ecm~=~\SI{65}{keV}$, which thus plays a key role in AGB nucleosynthesis. 
A footprint of this nucleosynthesis is the oxygen isotopic abundances observed in stellar spectra~\cite{Tsuji-2008} or in pre-solar meteoritic grains~\cite{Meyer-2008}. Moreover, the oxygen isotopic abundance ratios are a sensitive tracer of the mixing process in AGB stars. Observations can provide constraints for stellar models only if accurate thermonuclear reaction rates for the oxygen proton capture reactions are available over a wide region of stellar temperatures. At present only indirect measurements exist for the $\Ecm~=~\SI{65}{keV}$ resonance~\cite{Mak80,Landre89,Blackmon95} and it was deemed too weak to be measured even at LUNA with the setup described above. Its strength is presently evaluated as $\omega\gamma$~=~(16\,$\pm$\,3)\,peV from the known resonance strengths of \react{17}{O}{\pa}{14}{N}, \react{14}{N}{\ag}{18}{F} and the $\Gamma_\alpha$ from the \react{14}{N}{\alal}{}{} scattering reaction~\cite{Fox2005}.
Based on this value and assuming a target enrichment of  {90}\% and a detection efficiency of  {100}\%, the estimated reaction count rate is of the order of \SI{0.25}{counts/Coulomb}. In order to reach the required sensitivity and perform the first direct measurement of the \SI{65}{keV} resonance LUNA has designed a new shielding for the high efficiency 4$\pi$~BGO detector, as shown in Figure~\ref{fig2_17Opg}. The shielding is made of two layers: the inner one is a lead shielding and the external one consists of \SI{2}{inches} of borated polyethylene. The latter is crucial to absorb environmental neutrons and reduce backgrounds caused by neutron capture processes in the BGO detector~\cite{Boeltzig18-JPG}. Major efforts were devoted to investigate and reduce the beam induced background. The detection limit~\cite{Gilmore08-Book} was calculated assuming typical beam current of \SI{100}{\micro A} and a detection efficiency of  {48}\% in the region of interest, estimated by previous experimental campaigns. The environmental background with the complete setup was measured during August 2021 and it was taken into account for the LUNA sensitivity estimate. According to preliminary evaluations a resonance strength as low as $\omega\gamma\simeq$ 20\,peV can be detected.

LUNA has now the unique opportunity to determine experimentally such a low resonance strength by a direct measurement, with expected large impact on the \react{17}{O}{\pg}{18}{F} reaction rate and our understanding of AGB nucleosynthesis processes.

\subsubsection{\react{18}{O}{\pa}{15}{N}} % Carlo
The \react{18}{O}{\pa}{15}{N} reaction ($Q_{value}~=~\SI{3.98}{MeV}$) plays a central role in the production of the stable \nuc{15}{N}, \nuc{18}{O}, \nuc{19}{F} isotopes which can be used to trace the mixing and nucleosynthetic processes occurring inside AGB stars and novae. Of particular interest is the \nuc{18}{O}/\nuc{16}{O} isotopic abundance. Predictions of this abundance are at odds with observations in the Group-II silicate grains mentioned above, and dilution with solar matter is currently assumed as an explanation. Improved constraints on the \react{18}{O}{\pa}{15}{N} reaction would help investigate this issue.

\begin{figure}[h!]
\begin{center} 
    \includegraphics[width=0.7\textwidth]{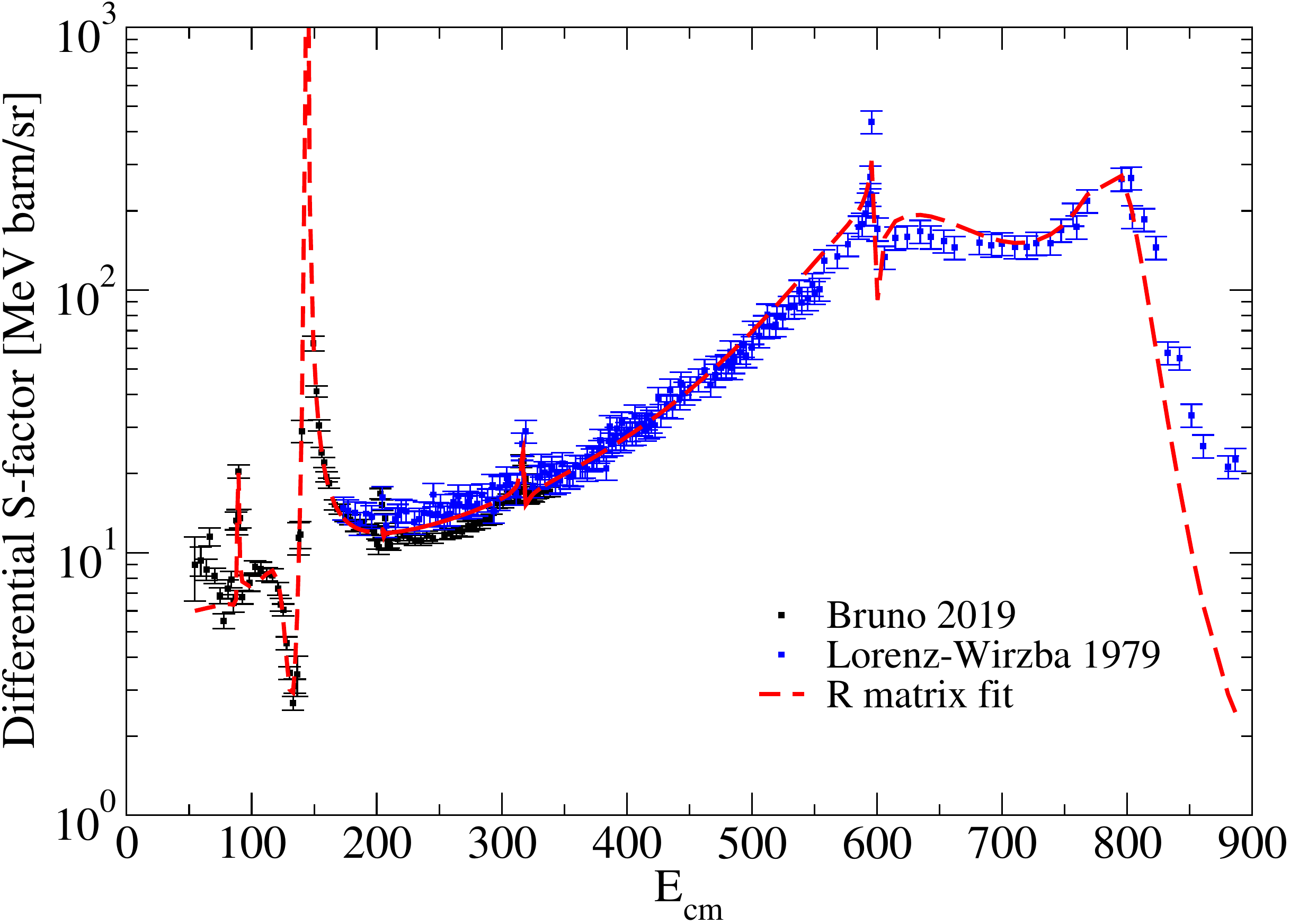}
    \caption{ $R$-matrix fit of the data acquired at LUNA on the \react{18}{O}{\pa}{15}{N} reaction, as well as data reported {in}~\cite{LW79}. Adapted from~\cite{Bruno2019}.} %MDPI: Please confirm if copyright permission is needed for this figure, please check all figures
    \label{fig:pa_rmatrix}
\end{center}
\end{figure}

At temperatures of astrophysical interest ($T$ = \SIrange{10}{1000}{MK}), the rate of this reaction is affected by a complex pattern of resonances. In particular, three interfering resonances with $\Jpi~=~1/2^+$ at $\Ecm~=~\num{143}$, \num{610} and \SI{800}{keV} dominate the rate (see Table~\ref{tab:18Opa}). The width and energy of the latter two resonances are controversial in the literature~\cite{Lacognata10}, and available data sets show tensions at energies $\Ecm \le \sim$$\SI{1}{MeV}$~\cite{Lacognata10,Mak78,LW79}. At LUNA, this reaction was investigated in the energy range $\Ep=\num{360}$ to $\SI{60}{keV}$, the lowest energy achieved to date, using the same setup employed for the \react{17}{O}{\pa}{14}{N} (Figure~\ref{fig:pa_chamber}). Data were analysed using a multi-channel $R$-matrix fit, as described in~\cite{Bruno2019}. The final fit is shown in Figure~\ref{fig:pa_rmatrix}, and numerical results are reported in Table~\ref{tab:18Opa}. In particular, the $\Ecm~=~\SI{90}{keV}$ resonance was found to be a factor \num{5} higher than previously found, and a new, previously unobserved state at $\Ecm~=~\SI{106}{keV}$ was required to reproduce the trend of data at low energies~\cite{Bruno2019}. These results significantly reduced the uncertainty in the stellar rate of the \react{18}{O}{\pa}{15}{N} reaction, and confirmed the need for dilution with solar matter to explain the abundances observed in oxygen-rich Group-II silicate grains using the model proposed in~\cite{Lugaro2017}.

\begin{table}[h!]
\caption{Partial widths and spin-parities of the resonances in the \react{18}{O}{\pa}{15}{N} reaction. Int.\ refers to the sign of the off-diagonal interference. See text for details.}
%\begin{center}
\setlength{\tabcolsep}{6.2mm}
\begin{tabular}{c c c c c}
\hline
\boldmath{\Ecm{} \textbf{[keV]}} & \boldmath{\Jpi{}}  & \boldmath{$\Gamma_p$\,\textbf{[eV]}} & \boldmath{$\Gamma_\mathrm{\alpha}$\,\textbf{[eV]}} & \textbf{Int.}  \\
\hline
89\,$\pm$\,0.3	&	3/2+	&	(797\,$\pm$\,57)\,10$^{-9}$	&	121\,$\pm$\,5	&	+	\\
106\,$\pm$\,3	&	1/2$-$	&	(120\,$\pm$\,10)\,10$^{-6}$	&	(86\,$\pm$\,1.6)\,10$^3$	&	+	\\
142.8\,$\pm$\,0.3	&	1/2+	&	(164\,$\pm$\,12)\,10$^{-3}$	&	150\,$\pm$\,1	&	+	\\
204.7\,$\pm$\,0.3	&	5/2+	&	(791\,$\pm$\,56)\,10$^{-6}$	&	12\,$\pm$\,1	&	+	\\
317.2\,$\pm$\,0.3	&	5/2+	&	(28\,$\pm$\,2)\,10$^{-3}$	&	(1.9\,$\pm$\,0.1)\,10$^3$ &	$-$	\\
597.6\,$\pm$\,0.3	&	3/2$-$	&	36\,$\pm$\,2	&	(2.5\,$\pm$\,0.1)\,10$^3$	&	+	\\
612.5\,$\pm$\,1.2	&	1/2+	&	(7.7\,$\pm$\,0.1)\,10$^3$	&	(163\,$\pm$\,1)\,10$^3$	&	$-$	\\
799.8\,$\pm$\,0.3	&	1/2+	&	(24.4\,$\pm$\,0.3)\,10$^3$	&	(26.1\,$\pm$\,0.3)\,10$^3$ &	+	\\
\hline
\end{tabular}
%\end{center}
\label{tab:18Opa}
\end{table}

\vspace{-9pt}

\subsubsection{\react{18}{O}{\pg}{19}{F}} % Axel

As a competing reaction to \react{18}{O}{\pa}{15}{N} discussed above, the \react{18}{O}{\pg}{19}{F} is of interest to understand the ratio of the two branches in stellar scenarios. At temperatures below 0.1 GK,  the reaction rate for the $\pg{}$ channel is determined by a low-energy resonance at $\Ep = \SI{151}{keV}$, direct capture and a resonance at $\Ep = \SI{22}{keV}$~\cite{Buckner2012}. The strength of another resonance at $\Ep = \SI{90}{keV}$ was debated, with a discrepancy between an experimental upper limit of \SI{7.8}{neV} ({90}\%~CL) by \cite{Buckner2012}, and an indirect determination by \cite{Fortune2013} of (0.70 $\pm$ 0.28)\,$\mu$eV. A resonance strength on the order of \SI{}{\micro eV} would result in a significant impact of the \SI{90}{keV} resonance on the stellar reaction rate around \SI{50}{MK}.

Underground measurements at LUNA-400 studied this reaction with Ta$_2$O$_5$ targets enriched in \nuc{18}{O}. To search for the $\Ep = \SI{90}{keV}$ resonance, the high-efficiency BGO detection setup was used in total absorption spectroscopy mode. This allowed us to exploit the low environmental background in the region of interest of the reaction ($Q_{value} = \SI{7.994}{MeV}$) in the underground laboratory. Target thickness and stability were monitored by scanning the narrow resonance at $\Ep = \SI{151}{keV}$ of the same reaction. Observations in this experiment resulted in a strength of (0.53\,$\pm$\,0.07)\,neV~\cite{Best2019}, compatible with the previous upper limit by \cite{Buckner2012}, and thus excluded a significant contribution of this resonance to the stellar reaction rate.

In a follow-up experiment at LUNA-400, the narrow resonances in \react{18}{O}{\pg}{19}{F} at higher energies were studied with a high-purity germanium detector, which was shielded by \SI{15}{cm} of lead and located at 55 degree to the beam axis. Revisiting the resonances at at $\Ep = \num{151}$, \num{215}, \num{274} and \SI{334}{keV} after the extensive work by \cite{Wiescher1980} resulted in the observation of numerous primary transitions with small branching ratios, which had not been previously observed in experiments on surface. Overall, the contributions of these resonances to the stellar reaction rate of \react{18}{O}{\pg}{19}{F} were confirmed \cite{Pantaleo2021}.

\subsubsection{\react{12}{C}{\pg}{13}{N} and \react{13}{C}{\pg}{14}{N}}
% Jakub, Axel

\begin{figure}[h!]
\begin{center}    
    \vspace*{-0.7cm}
    \hspace*{-1.35cm}
    \includegraphics[width=\textwidth]{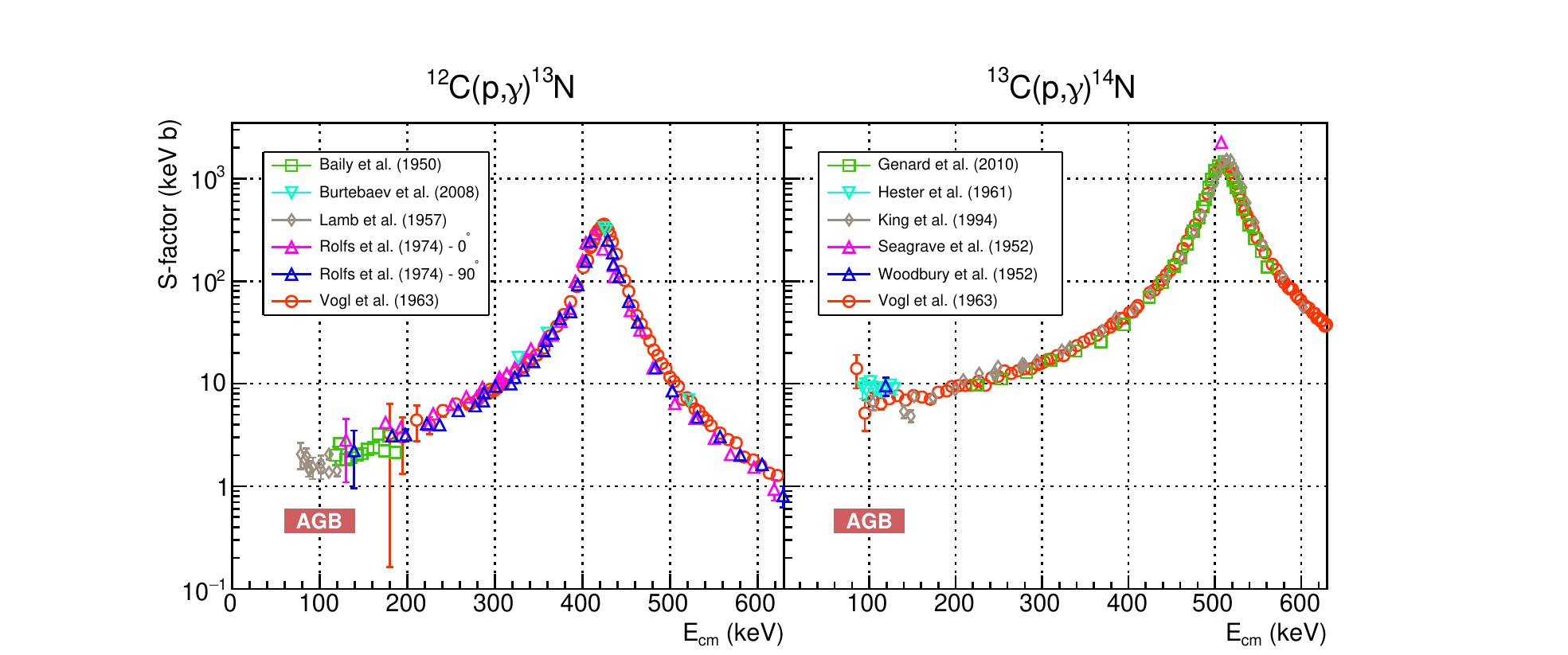}
    \caption{State of the art for the \react{12}{C}{\pg}{13}{N} and \react{13}{C}{\pg}{14}{N} reactions. For the latter, only the transition 
    to the ground state is considered.}
    \label{fig:cpg}
\end{center}
\end{figure}

Both \react{12}{C}{\pg}{13}{N} and 
\react{13}{C}{\pg}{14}{N} are the first two reactions of the
CNO cycle and are of particular interest since they determine the 
isotopic abundance of carbon inside the hydrogen burning regions. In fact,
the \nuc{12}{C}/\nuc{13}{C} ratio in the stellar surface layers can be 
used as a tracer of stellar evolution since it is readily 
derived from stellar spectra~\cite{Savage2008}. Changes in the isotopic ratio occur when stars evolve from the main sequence towards the Red Giant branch. In particular, during the 
AGB helium flashes, the hydrogen burning shell temporarily stops 
burning. As a consequence, the convective envelope penetrates 
into the C-rich and H-exhausted layers, bringing the freshly 
synthesized products to the surface. This leads to an increase in the
\nuc{12}{C}/\nuc{13}{C} ratio on the stellar surface. Currently the success in reproducing \nuc{12}{C}/\nuc{13}{C} ratios appears to indicate that AGB stars experience extra mixing episodes~\cite{Palmerini2011}. Furthermore, discrepancies were found in the recent analysis of the presolar SiC grains ~\cite{Liu2021}, where it is evidenced that the discrepancy could be solved by extra mixing caused by the magnetic buoyancy~\cite{Palmerini2017}. Hence, further insight into complex AGB mixing events can be obtained by constraining the isotopic ratio predictions. Since the ratio of carbon isotopes is highly sensitive to the adopted rates of proton capture reactions, a precise measurement of the magnitude of these reactions is needed to obtain reliable predictions. Numerous past experiments studied
the cross-section of both \react{12}{C}{\pg}{13}{N}~\cite{Baily1950,Lamb1957,Rolfs1974,Burtebaev2008,Vogl1963}, and \react{13}{C}{\pg}{14}{N}~\cite{Vogl1963,King1994,Seagrave1952,Woodbury1952,Hester1961,Genard2010}. Nonetheless, as shown in Figure~\ref{fig:cpg}, data at the lowest energies are subject to significant scatter and large uncertainties, affecting extrapolations of the $S$-factor.

Recently, \react{12}{C}{\pg}{13}{N} and \react{13}{C}{\pg}{14}{N} were
measured at LUNA with complementary experiments, all using solid carbon targets. By using different experimental techniques, measurements can be cross-checked and systematic uncertainties reduced.
In the first campaign, both reactions were studied using an HPGe detector in close geometry. This permitted a precise spectroscopic
study to extract the cross-section at energies as low as $\Ecm = \SI{80}{keV}$. Data analysis for this experiment is at an advanced stage. The main procedure for both the $S$-factor extraction and the target monitoring was the improved application of the $\gamma$-shape analysis~\cite{Ciani2020}. A second experimental campaign was performed with the BGO detector, which permitted to measure the reactions at even lower 
energies. In the case of \react{13}{C}{\pg}{14}{N}, the high  $Q$-value ($\sim$$\SI{7.5}{MeV}$) made the total absorption spectroscopy particularly advantageous, since the sum  $\gamma$-peak lies in an almost background free region of the spectra. For \react{12}{C}{\pg}{13}{N}, with its smaller $Q$-value ($\sim$$\SI{1.9}{MeV}$), intrinsic backgrounds of the detector~\cite{Boeltzig18-JPG} are a limiting factor, and a different approach had to be found. Given that \nuc{13}{N} is $\beta^+$ unstable with a half-life of about \SI{10}{min}, it was possible to detect the \SI{511}{keV} $\gamma$-rays from positron-electron annihilation in coincidence in two opposite crystals of the detector. This distinct experimental signature helped to significantly reduce the background counting rate in the detector. Thanks to the experimental setup, the analysis of the \SI{511}{keV} rate can be carried out not 
only by the use of the post-irradiation, but also by using the irradiation 
period itself. The analysis of the data acquired with the BGO detector is ongoing, but preliminary results looks promising and in agreement with those obtained in the first HPGe campaign.

%%%%%%%%%%%%%%%%%%%%%%%%%%%%%%%%%%%%%%%%%%

\subsection{NeNa Cycle}
At the base of the convective envelope of massive AGB stars, temperatures up to $\sim$$\num{0.1}$~GK activate H-burning cycles involving heavier elements than CNO, namely the NeNa and MgAl cycles. The onset of deep convective motions and the so-called hot bottom burning (HBB) process~\cite{Boothroyd-95ApJ} brings to the surface the material processed in the H-burning shell~\cite{Herwig2005}. Therefore, interpretation of the observed abundances requires a detailed knowledge of the reaction network involved.

Over the last decade, LUNA has started to explore many of the reactions of the NeNa and MgAl cycles. The next paragraphs highlight some key results and ongoing experiments.

\subsection{\react{22}{Ne}{\pg}{23}{Na}}
\label{22ne_pg_23na} % Federico, Denise

Within the NeNa cycle, the \react{22}{Ne}{\pg}{23}{Na} reaction links \nuc{22}{Ne} to \nuc{23}{Na}. The former is  the third most abundant nuclide produced in stellar helium burning~\cite{Buchmann06} and an important neutron source for the astrophysical $s$-process in AGB stars and in massive stars~\cite{Kaeppeler2011}. The latter is the only stable isotope of sodium. This reaction can influence one of the most puzzling recent astronomical observations, the anti-correlation of oxygen and sodium abundances observed in globular clusters~\cite{Carretta2009, Gratton2012}.

Among the reactions of the NeNa cycle the \react{22}{Ne}{\pg}{23}{Na} has been for several years the one with the most uncertain reaction rate, heavily affecting models seeking to reproduce the oxygen-sodium anti-correlation~\cite{Ventura2018} as well as the nucleosynthesis up to \nuc{24}{Mg} isotopes.
At the temperatures of interest for the HBB in AGB stars, $\num{0.07} \le T \le \SI{0.1}{GK}$, the astrophysical reaction rate is dominated by a number of resonances, whose strength must be known precisely. Long before the recent LUNA campaigns, one direct measurement was reported, providing only upper limits~\cite{Goerres82-NPA}. Results of indirect measurements were also available, but the resonance strengths were based on uncertain spin parity assignments or spectroscopic factors.
As a result, a discrepancy of a factor of 1000 existed between the recommended rates from the NACRE compilation~\cite{ANGULO1999}, on the one hand, and the evaluation by STARLIB~\cite{Sallaska13-ApJSS}, on the other hand. The discrepancy was mainly due to the treatment of the resonances below \SI{400}{keV}. 

Two experimental campaigns were carried out at LUNA. The main goal of the first campaign was the direct study of the resonances mentioned above, exploiting the deep underground low-background environment.
The experimental setup consisted of a windowless gas target chamber filled with \SI{1.5}{mbar} \nuc{22}{Ne} gas (isotopic enrichment  99.9\%), recirculated and purified by a chemical getter and two large, collimated high-purity germanium detectors, at 55° (Ge55) and 90° (Ge90), respectively, to the beam axis~\cite{Cavanna2014}. The scattering chamber and the detectors were enclosed in a copper and lead shielding, to reduce the background caused by natural radioactivity in the laboratory Figure~\ref{fig:22Ne(p,g)23Na_setup}b.
To further reduce the background due to \nuc{222}{Rn}, the whole shielding was enclosed in an acrylic box flushed with nitrogen vapour.
\begin{figure}[h!]
\begin{center}
 %   \centering
    \subfigure[]
       { \includegraphics[width=0.48\textwidth]{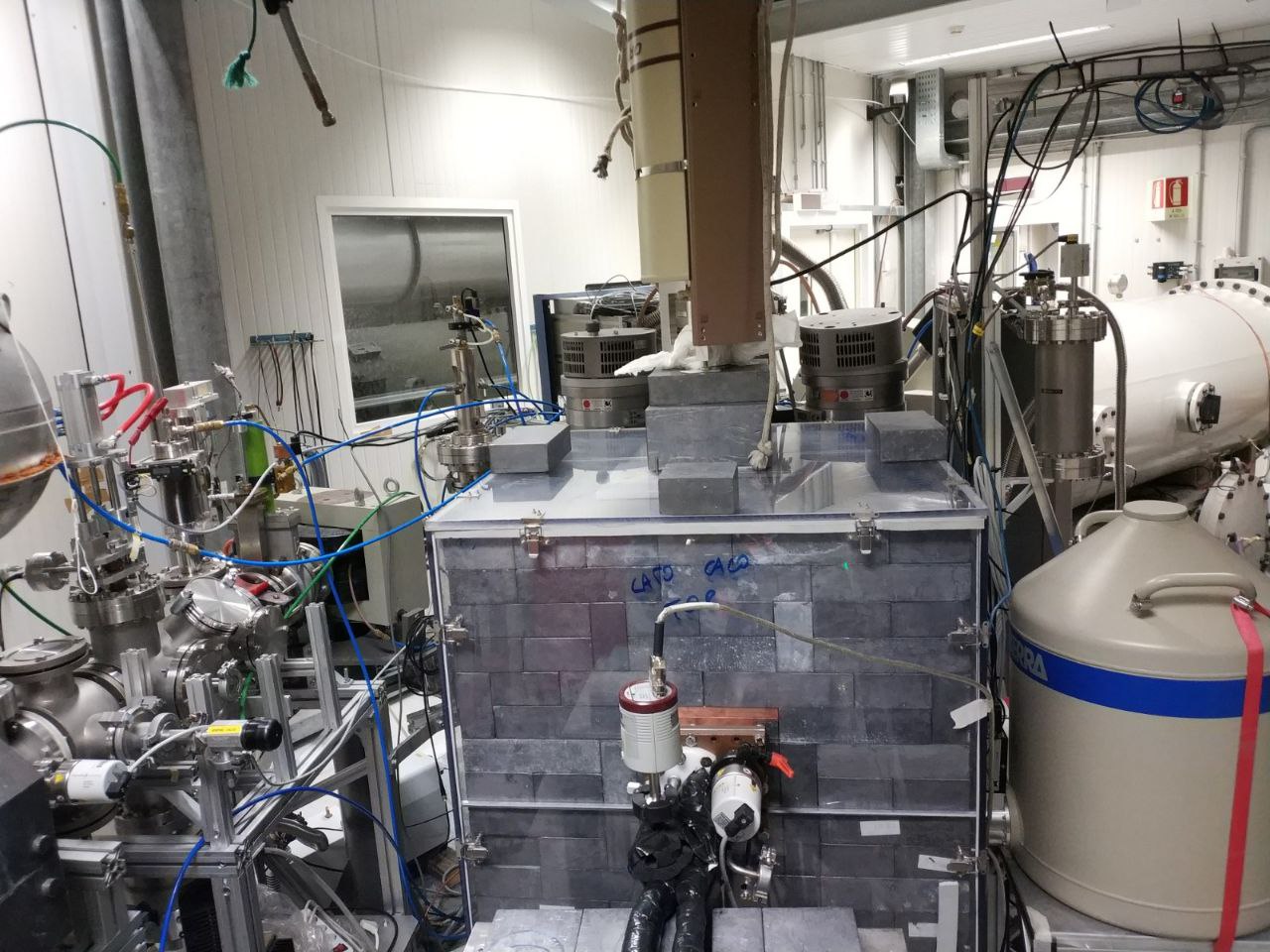}}
      %  \caption{Setup used during the HPGe campaign. The Dewars of the two HPGe detectors are visible, close to the lead shielding and the radon box.}
        \label{fig:22Ne(p,g)23Na_BGOsetup}
   % \end{subfigure}
 %   \hfill
   \subfigure[]
        { \includegraphics[width=0.48\textwidth]{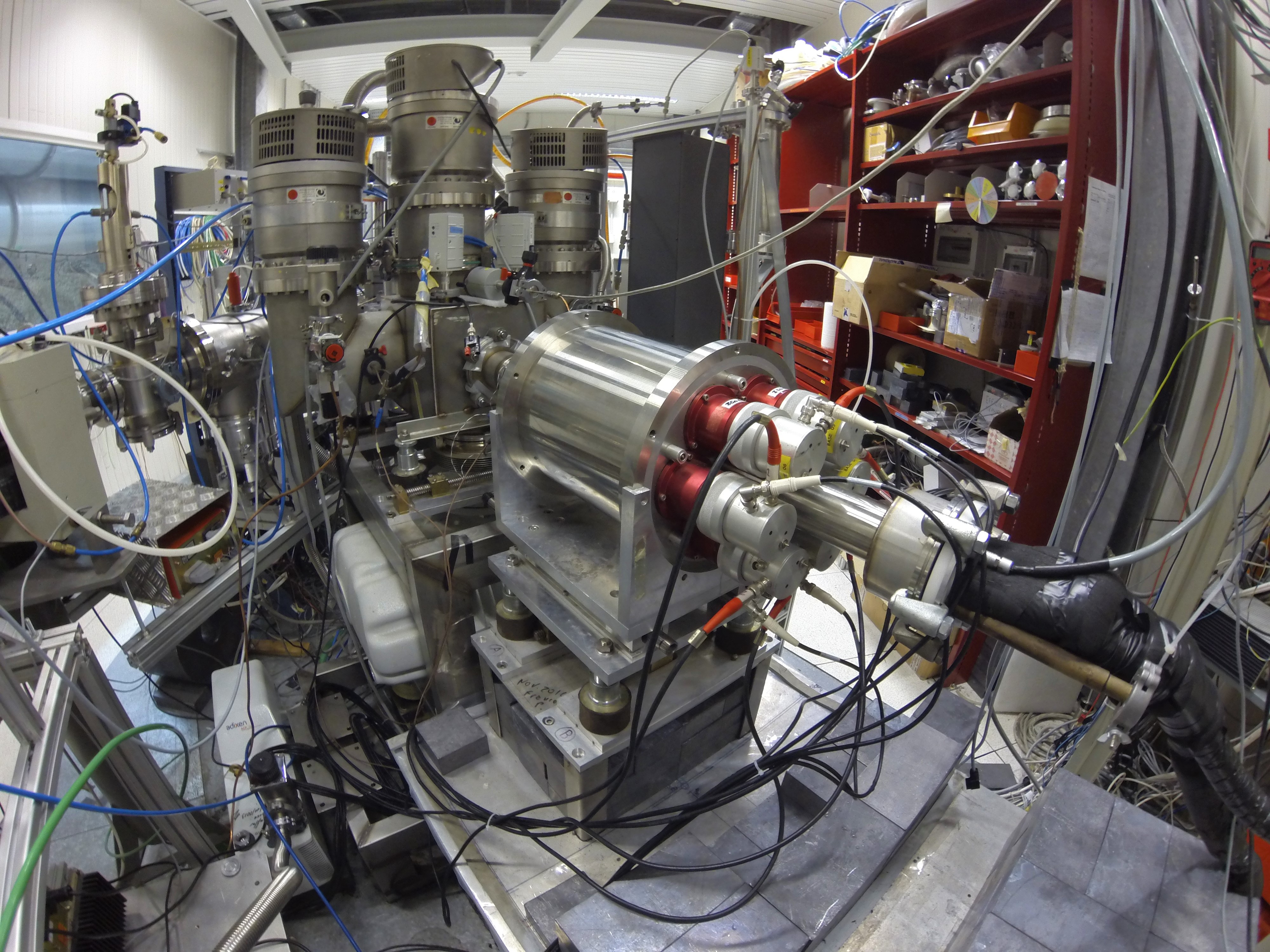}}
         %\caption{Setup used during the BGO campaign. The beamline is passing through the segmented BGO detector, that is standing unshielded.}
         \label{fig:22Ne(p,g)23Na_HPGeSetup}
     %\end{subfigure}

     \caption{The two experimental setups used in the LUNA campaigns on the \react{22}{Ne}{\pg}{23}{Na} reaction. (\textbf{a}) Setup used during the HPGe campaign. The Dewars of the two HPGe detectors are visible, close to the lead shielding and the radon box. (\textbf{b}) Setup used during the BGO campaign. The beamline is passing through the segmented BGO detector, that is standing unshielded.   \label{fig:22Ne(p,g)23Na_setup}}
\end{center}
\end{figure}

Three new resonances were observed for the first time, at proton energies of (156.2\,$\pm$\,0.7), (189.5\,$\pm$\,0.7) and (259.7\,$\pm$\,0.6)\,keV~\cite{Cavanna2015}.
In addition, more precise \nuc{23}{Na} excitation energies corresponding to the new resonances were determined. For other three resonances, at \num{71}, \num{105} and \SI{215}{keV}, new upper limits to the strengths were obtained, see Table~\ref{tab:wg_TRR}.
The impact of the new LUNA reaction rate was studied focusing on the chemical ejecta of thermally pulsing asymptotic giant branch (TP-AGB) stars that experience hot-bottom burning~\cite{slemer2016}, reproducing observations of the O–Na anticorrelation in galactic globular clusters~\cite{Carretta2009}.

\begin{table}[h!]
\caption{{Strengths} of low energy resonances in the \react{22}{Ne}{\pg}{23}{Na} reaction. Upper limits are given at  {90}\% confidence level.}  
\label{tab:wg_TRR}
\resizebox{\columnwidth}{!}{
\begin{tabular}{ c   c   c   c   c}
\hline
\multicolumn{1}{c }{\boldmath\textbf{$E_{\rm p}^{\rm res}$}} & \multicolumn{4}{c}{\textbf{Strength} \boldmath{$\omega\gamma$}\,\textbf{[eV]}} \\%& \multicolumn{1}{c}{Screening} \\
 \textbf{[keV]}   & \textbf{Iliadis  {et al.}}   & \textbf{LUNA-HPGe}                         & \textbf{TUNL}              & \textbf{LUNA-BGO} \\%& \multicolumn{1}{c|}{Adopted} & \multicolumn{1}{c}{enhancement} \\
            &~\textbf{\cite{Iliadis10-NPA}}      &~\textbf{\cite{Cavanna2015}}    &~\textbf{\cite{Kelly2017}}  & \textbf{\cite{Ferraro2018-PRL}}  \\%& & \multicolumn{1}{c}{factor $f$} \\
\hline
 
% 29 & - & $\leq$\,2.6\,$\times$\,10$^{-25}$\,\cite{Iliadis10-NPA} & - & - \\
37      & (3.1\,$\pm$\,1.2)\,$\times$\,10$^{-15}$       & -                                         & - & - \\ 
71      & -                                             & $\leq$1.5\,$\times$\,10$^{-9}$          & - & $\leq$6\,$\times$\,10$^{-11}$ \\
105     & -                                             & $\leq$7.6\,$\times$\,10$^{-9}$          & - & $\leq$7\,$\times$\,10$^{-11}$ \\
156.2   & (9.2\,$\pm$\,3.0)\,$\times$\,10$^{-9}$        & (1.8\,$\pm$\,0.2)\,$\times$\,10$^{-7}$    & (2.0\,$\pm$\,0.4)\,$\times$\,10$^{-7}$ & (2.2\,$\pm$\,0.2)\,$\times$\,10$^{-7}$ \\
189.5   & $\leq$2.6\,$\times$\,10$^{-6}$              & (2.2\,$\pm$\,0.2)\,$\times$\,10$^{-6}$    & (2.3\,$\pm$\,0.3)\,$\times$\,10$^{-6}$ & (2.7\,$\pm$\,0.2)\,$\times$\,10$^{-6}$ \\
215     & -                                             & $\leq$2.8\,$\times$\,10$^{-8}$          & - & - \\
259.7   & $\leq$1.3\,$\times$\,10$^{-7}$              & (8.2\,$\pm$\,0.7)\,$\times$\,10$^{-6}$    & - & (9.7\,$\pm$\,0.7)\,$\times$\,10$^{-6}$ \\
% 291 & $\leq$\,2.2\,$\times$\,10$^{-6}$\,\cite{Goerres82-NPA} & - & - & - \\
% 323 & $\leq$\,2.2\,$\times$\,10$^{-6}$\,\cite{Goerres82-NPA} & - & - & - \\
% 334 & $\leq$\,3.0\,$\times$\,10$^{-6}$\,\cite{Goerres82-NPA} & - & - & - \\
% 369 & - & $\leq$\,6.0\,$\times$\,10$^{-4}$\,\cite{Hale01-PRC} & - & - \\
% 394 & - & $\leq$\,6.0\,$\times$\,10$^{-4}$\,\cite{Hale01-PRC} & - & - \\
% 436 & 0.079 $\pm$ 0.006~\cite{Depalo15-PRC} & - & - & - \\
% 479 & 0.594 $\pm$ 0.038~\cite{Depalo15-PRC} & - & - & - \\
% 638.5 & 2.45 $\pm$ 0.18~\cite{Depalo15-PRC} & - & - & - \\
% 661 & 0.032 $\pm$ 0.015~\cite{Depalo15-PRC} & - & - & - \\
\hline          
\end{tabular}
}
 
\end{table}%

A high efficiency measurement was then performed to investigate the still undetected resonances {71} and {105} {keV}, as well as the non resonant contribution. The second experimental campaign was carried out using the same gas target, and a new scattering chamber surrounded by the LUNA 4$\pi$ BGO detector. In order to quantify and subtract the beam induced background contribution, each run with \nuc{22}{Ne} was alternated to a run with inert Ar gas. The run in Ar was used to subtract the beam-induced background from the measurements in \nuc{22}{Ne}~\cite{Ferraro2018-EPJ}. 
The resonance strengths of the three resonances at $\Ep=\num{156.2}$, \num{189.5}, and \SI{259.7}{keV} measured during the new campaign is consistent within $2\sigma$ with the measurements obtained during the first campaign.
New more stringent upper limits were calculated for the \num{71} and \SI{105}{keV} resonance, which now play a negligible role in the reaction rate, see Table~\ref{tab:wg_TRR}. 
The non-resonant component of the cross section was measured at four energies: \num{183.8}, \num{200.8}, \num{245.9} and \SI{306.1}{keV}. These energies were chosen to avoid any sizeable contribution of the known resonances to the measurement of the non-resonant component.

The uncertainty of the reaction rate was greatly reduced and the discrepancy between the most frequently used rates~\cite{ANGULO1999, Heil2008} has now been overcome at temperatures of interest for the HBB.

It is worth noting that an additional direct measurement at a surface facility of the resonances at 436, 479, 639, 661, and 1279 keV~\cite{Depalo15-PRC} improved the knowledge of the corresponding resonance strengths and contributed to the new, improved calculation of the rate. The new calculated reaction rate significantly increases at temperatures $\num{0.08} \leq T \leq \SI{0.3}{GK}$ relevant to shell H-burning in AGB stars and classical novae explosions, as shown in~\cite{Cavanna2015,Depalo2016}.

The new LUNA reaction rate provides a precise tool to be used for future studies addressing nucleosynthesis in AGB stars.
\subsubsection{\react{20}{Ne}{\pg}{21}{Na}} %Eliana, Alessandro

After the precise measurement of the \react{22}{Ne}{\pg}{23}{Na} reaction at LUNA, the reaction \react{20}{Ne}{\pg}{21}{Na} is now the major source of uncertainty on the reaction rate network of the NeNa cycle.
The \react{20}{Ne}{\pg}{21}{Na} reaction may take place in several stellar environments such as  H shell-burning in red giants stars and asymptotic giant branch stars, classical novae explosions, and core H-burning in massive stars, where temperatures can reach $T=\SI{0.05}{GK}$~\cite{Marion57-ApJ}. The proton capture on \nuc{20}{Ne} nuclei at low energies could help explaining the observed oxygen-sodium (O--Na) abundance anticorrelation in globular cluster (GC) member stars~\cite{Gratton04-ARAA}. This is due to the fact that the \react{20}{Ne}{\pg}{21}{Na} reaction is the slowest reaction of the NeNa cycle, therefore it sets the speed of the entire cycle and serves as a bottleneck for the production of \nuc{22}{Na} and  \nuc{22}{Ne}. 
Moreover, as pointed out in~\cite{Renzini_2015} difficulties exist in modelling the O--Na anticorrelation  without sodium destruction. A reduction of about 2.5 in the \react{23}{Na}{\pa}{20}{Ne} reaction cross section might limit the sodium destruction and solve the observed abundances~\cite{Ventura06-AA}. A recent work~\cite{Kelly2017} showed that the exit (\react{23}{Na}{\pa}{20}{Ne}) and entrance (\react{20}{Ne}{\pg}{21}{Na}) reactions for the NeNa cycle are the least known and therefore a better knowledge of both can fix the abundance issues.

Below \SI{0.1}{GK}, the \react{20}{Ne}{\pg}{21}{Na} reaction rate is dominated by the high energy tail of a sub-threshold state at $\Ecm$~=~\SI{-6.7}{keV}. For temperatures between \SI{0.1}{GK} and \SI{1}{GK} the rate is influenced by the direct capture component and the narrow resonance at $\Ecm$~=~\SI{366}{keV}, both accessible with the LUNA~\SI{400}{kV} accelerator.

To date, the \react{20}{Ne}{\pg}{21}{Na} reaction has been studied using both direct~\cite{Rolfs75-NPA} and indirect~\cite{Mukhamedzhanov} approaches. The direct capture cross section was studied down to \SI{370}{keV} proton energy in~\cite{Rolfs75-NPA}. The same paper presents a measurement of the $\Ecm$~=~\SI{366}{keV} resonance strength, reporting a $\omega\gamma$\,=\,(0.11\,$\pm$\,0.02)\,meV~\cite{Rolfs75-NPA}.
A more recent study reports a measurement of the \react{20}{Ne}{\pg}{21}{Na} reaction over a wide energy range ($\Ep$~=~\SIrange{0.5}{2.0}{MeV}) and provides an $R$-matrix analysis to determine the contributions of the sub-threshold resonance and the direct capture to the total cross Section \cite{Lyons2018}. The $R$-matrix fit was used to extrapolate the S-factor down to low energies. They calculated the total reaction rate with the new data and found an overall $\sim${20}\% reduction  with respect to previous data.  

The LUNA collaboration aims to measure both the \SI{366}{keV} resonance and the direct capture below \SI{400}{keV}, where no experimental data exist. 
The first campaigns for the \SI{366}{keV} resonance were completed using an experimental setup similar to the one used for the study of the \react{22}{Ne}{\pg}{23}{Na} reaction, germanium phase~\cite{Cavanna2015, Depalo2016}. Compared to the \react{22}{Ne}{\pg}{23}{Na} setup, for the ongoing measurement the inner collimators have been removed. In addition, the physical position of the two germanium detectors allows to experimentally parametrise the energy straggling of the proton beam passing through the neon gas, and correct its effect using a different approach with respect to the one given in~\cite{Bemmerer18-EPL}. The energy straggling affects the position inside the chamber where a nuclear reaction takes place. In the particular case of the resonant reactions, if the energy straggling is negligible, the resonance takes place in a thin slice of the target. Instead, for a large beam spread the resonance is populated in a thicker slice of the target, causing a reduction of the maximum resonance yield. The data analysis for the \SI{366}{keV} resonance is still ongoing and preliminary results are in agreement with~\cite{Rolfs75-NPA}. The same experimental setup will be used to study the direct capture component down to \SI{250}{keV}.

\subsubsection{\react{23}{Na}{\pg}{24}{Mg}} % Axel

The reaction \react{23}{Na}{\pg}{24}{Mg} provides a transition from the NeNa cycle to the MgAl cycle, competing with the \react{23}{Na}{\pa}{20}{Ne} reaction as part of the NeNa cycle. 
The impact of a resonance at $\Ep \simeq \SI{140}{keV}$ on the reaction rate has been studied in the past. Upper limits on the strength of this resonance were determined by \cite{Goerres1989} ($\leq$$\SI{0.5}{\micro eV}$) and later improved by experiments of \cite{Rowland2004} ($\leq$$\SI{0.15}{\micro eV}$) and \cite{Cesaratto2013} ($\leq$$\SI{5.17}{neV}$). The latter measurement reports an excess of observed events over background, not quite reaching statistical significance for a clear observation of this resonance, but possibly indicating its important role for the reaction rate around \mbox{$T = 0.07$ GK}.

With a $Q$-value of \SI{11693}{keV}, \react{23}{Na}{\pg}{24}{Mg} is an ideal case for total absorption spectroscopy in an underground environment, thanks to the effectively suppressed cosmic background at these energies. The search for the $\Ep$ =~\SI{140}{keV} resonance has been continued at LUNA-400, using the BGO summing detector. With the strongly reduced environmental backgrounds~\cite{Boeltzig18-JPG}, beam-induced backgrounds from reactions with larger $Q$-values, such as \react{11}{B}{\pg}{12}{C}, were critical in this measurement. After careful data analysis, a signal was observed leading to a resonance strength of (1.46\,$+$\,0.58\,$-$\,0.53)\,$\mathrm{neV}$~\cite{Boeltzig2019}. This improved knowledge of the resonance strength significantly reduced the uncertainty of the reaction rate in the temperature region where this resonance is dominant.

A recent measurement by \cite{Marshall2021} revised the resonance energy, with a consequent reduction of the reaction rate up to a factor of 2 around \SI{70}{MK}. Current experimental efforts at TUNL aim to further reduce the uncertainty of this energy determination.

Two stronger, narrow resonances at $\Ep=\SI{251}{keV}$ and $\SI{309}{keV}$ were also studied at LUNA-400, using a shielded HPGe detector in the same configuration as described for the \react{18}{O}{\pg}{19}{F} campaign. Branching ratios for the resonance at \SI{251}{keV} were established, and the strength of this narrow resonance determined with smaller uncertainty than previously known~\cite{Boeltzig2019}, reducing the reaction rate uncertainty around $T=\num{0.1}$ GK.

An experimental determination of the direct capture component, currently determined with a relative uncertainty of \SI{40}\%~\cite{Hale2004}, and the properties of the \SI{140}{keV} resonance are still needed. Measurements underground and on surface will be required to improve on this~knowledge.

%%%%%%%%%%%%%%%%%%%%%%%%%%%%%%%%%%%%%%%%%%
\section{{Neutron Sources for the \boldmath{$s$}-Process}}

Half of the elements heavier than iron are synthesized in the so called $s$-process, which consists of series of neutron capture reactions starting from Fe-group nuclei and $\beta$-decays along the valley of stability, reaching \nuc{209}{Bi} as the most massive stable nucleus. 
To reproduce the observed abundances of $s$-process elements, two sub-processes are defined according to the characteristic neutron density and metallicity of the stellar environment. 
These are so called weak- and main- components of the $s$-process~\cite{Straniero2006,Pignatari2010} responsible for the production of the elements with A~$\leq$~90 and 90~$<$~A~$\leq$~204, respectively. In current stellar models, Thermally Pulsing Low-mass Asymptotic Giant Branch (TP-AGB) stars and massive stars are considered to be the major production sites of $s$-process elements. The neutrons for the main $s$-process are predominantly supplied by the \react{13}{C}{\an}{16}{O} reaction, while \react{22}{Ne}{\an}{25}{Mg} provides most of the neutrons for the weak $s$-process~\cite{Busso01-APJ}. Although s-process nucleosynthesis is affected by the stellar structure, composition and mixing phenomena (see for example~\cite{Busso2021} and references therein), the knowledge of the cross sections of the two main neutron sources can still help improving the precision of predicted elemental abundances. 

Studies related to both reactions have been performed at LUNA-400 and are part of the scientific program of LUNA-MV, as illustrated in more details in the next paragraph.

\subsection{{\react{13}{C}{\an}{16}{O}}} % Giovanni, David, Laszlo
%MDPI: There is no Section 4.1, and we revised section numbers, please confirm -> OK
It is commonly believed that recursive mixing episodes in TP-AGB stars lead to the formation of a so-called \nuc{13}{C}-pocket through the reaction sequence \react{12}{C}{\pg}{}{}\react{13}{N}{(\beta^+ \nu)}{13}{C}~\mbox{\cite{Straniero2006,gallino1998}}. During the period between two TPs, the temperature attains about \SI{90}{MK} and \nuc{13}{C} starts capturing $\alpha$ particles. As a consequence neutrons are released through the reaction \react{13}{C}{\an}{16}{O}. This process provides a relatively slow neutron flux ($\approx$$10^{7}$\,neutrons/s/cm$^2$) for about $10^4$\,years each time. Starting from seed nuclei in the iron region, this neutron flux allows to build up heavy elements along the stability valley~\cite{Kaeppeler2011}. 

In order to constrain this important nucleosynthesis process, the cross section of the \react{13}{C}{\an}{16}{O} neutron source needs to be known in the astrophysical energy window around $\Ecm$~=~\SIrange{150}{230}{keV}. Several papers report direct~\cite{Brune1993, Drotleff1993, Heil2008} and indirect~\cite{Guo2012, Avila2015, Trippella2017} experiments to investigate the behavior of the \react{13}{C}{\an}{16}{O} reaction. However, experimental data at the lowest measured energies are affected by large uncertainties~\cite{Drotleff1993, Heil2008}. As a consequence, the extrapolation of the astrophysical S-factor into the Gamow window is challenging, especially due to the contested effect of a resonance located near the threshold~\cite{Cristallo2018, DeBoer2020}.

\begin{figure}[h!]
\begin{center} 
%    \vspace*{-0.7cm}
%    \hspace*{-1.35cm}
    \includegraphics[width=0.45\textwidth]{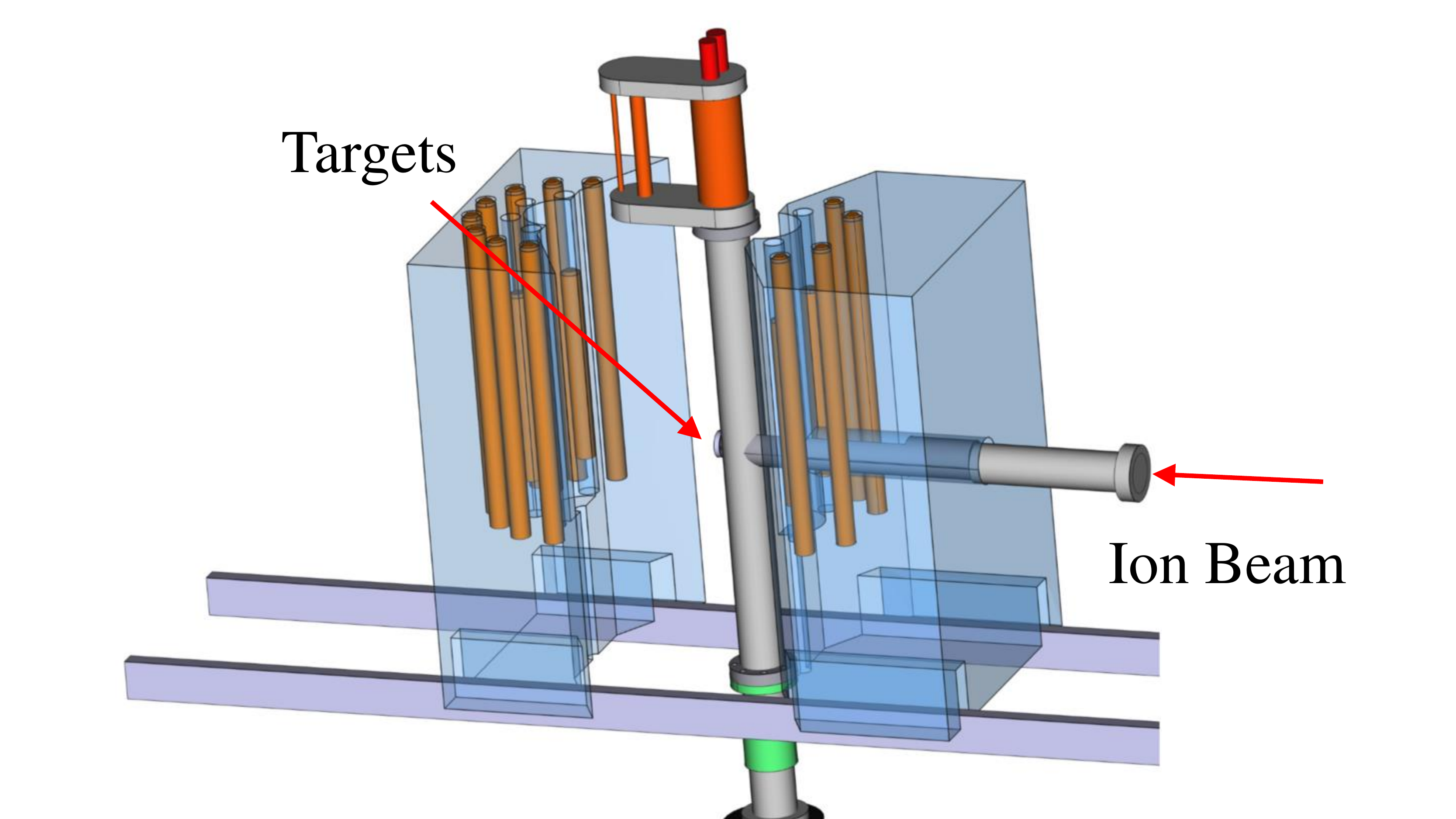}
    \includegraphics[width=0.45\textwidth]{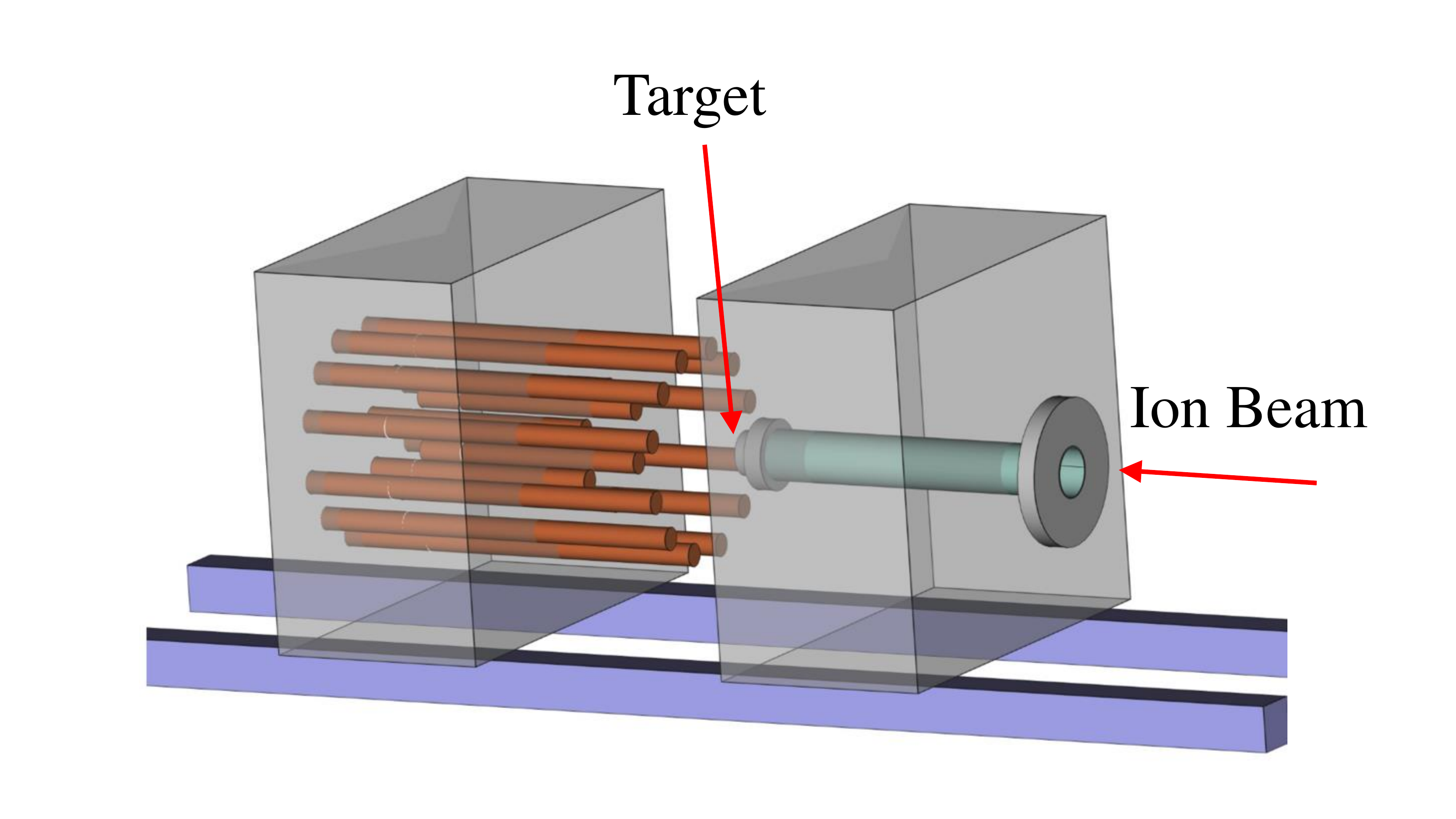}

   \caption{Vertical (left panel) and horizontal (right panel) setup of the LUNA neutron detector array. Orange tubes and transparent boxes represent the \nuc{3}{He} neutron counters and polyethylene moderators, respectively. Adapted {from}~\cite{Csedreki2021}.}
          \label{fig:CAN_setup} %MDPI: Please confirm if copyright permission is needed for this figure, please check all figures
\end{center}
\end{figure}

For the study of the \react{13}{C}{\an}{16}{O} reaction, the LUNA collaboration designed and developed a neutron detector array~\cite{Csedreki2021}. The experimental setup is based on eighteen \nuc{3}{He} counters with low intrinsic background settled in two concentric rings (6 in the inner ring, 12 in the outer ring) around the target chamber and embedded in a polyethylene moderator.
To measure the very low cross section of the \react{13}{C}{\an}{16}{O} reaction, the neutron detection efficiency, target handling, and active target cooling must be optimized. Therefore, two detector geometries with vertically and horizontally arranged counters were designed based on extensive GEANT4 simulations (Figure~\ref{fig:CAN_setup}).

%\begin{figure}
%    \centering
%    \begin{subfigure} 
%        \centering
%        \includegraphics[width=\textwidth]{Figures/CAN_Fig_a.pdf}
%    \end{subfigure}
%    \hfill
%    \begin{subfigure} 
%         \centering
%         \includegraphics[width=\textwidth]{Figures/CAN_Fig_b.pdf}
%         \end{subfigure}
%         \caption{(Adapted from~\cite{Csedreki2021}) Vertical (left panel) and horizontal (right panel) setup of the LUNA neutron detector array. Orange tubes and transparent boxes represent the \nuc{3}{He} neutron counters and polyethylene moderators, respectively.}
%          \label{fig:CAN_setup}
% \end{figure}

The location of the experiment in the LNGS laboratory, the use of \nuc{3}{He} counters with stainless steel housing and a post processing Pulse Shape Discrimination (PSD) technique~\cite{Balibrea2018} allowed an overall background reduction to 1 count/h in the entire setup, two orders of magnitude lower than previous experiments performed in surface laboratories.
The absolute neutron detection efficiency of the setup was determined using the \react{51}{V}{\pn}{51}{Cr} reaction and an AmBe radioactive source, and completed with a Geant4 simulation. \nuc{13}{C} targets used during the measurement at LUNA were produced by evaporating isotopically enriched (at  {99}\%) \nuc{13}{C} powder on tantalum backings.
They were characterized immediately after the production at the Tandem accelerator in ATOMKI, using the \SI{1.764}{MeV} narrow resonance of the \react{13}{C}{\pg}{14}{N} reaction. During the measurement at LUNA, the target stability was monitored by applying a line shape analysis technique to the ground state transition of the \react{13}{C}{\pg}{14}{N}~\cite{Ciani2020}. The procedure described above allowed us to measure the \react{13}{C}{\an}{16}{O} cross section in $\Ea$~=~\SIrange{305}{400}{keV} reaching an overall uncertainty lower than  {20}\% for all the data points~\cite{Ciani2021}.

\begin{figure}[h!]
\begin{center}    
    \includegraphics[width=0.9\textwidth]{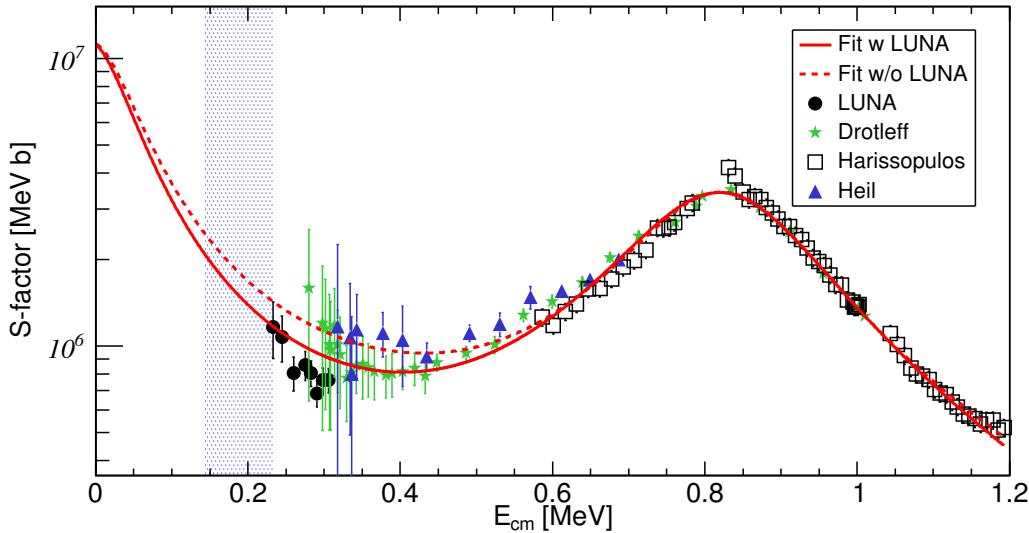}
    \caption{ R matrix extrapolation towards low energies including (solid red line) and excluding (dashed red line) the new LUNA data (black dots). For the R matrix evaluation also Harissopoulos, Drotleff and Heil data have been considered. The blue rectangle represents the Gamow peak range. Adapted {from}~\cite{Ciani2021}.} %MDPI: Please confirm if copyright permission is needed for this figure, please check all figures
    \label{fig:sfactor}
\end{center}
\end{figure}

The new data extended into the $s$-process Gamow peak, nevertheless a low energy extrapolation is still necessary to cover the entire astrophysical energy range. For this purpose a fit was performed with the AZURE2~\cite{Azuma2010} code in which all the relevant broad states were considered. Two high energy poles were added to account for the higher energy resonances effects. For the experimental data, in addition to the LUNA results, also higher energy data set (up to \SI{1.2}{MeV}) from the literature were included~\cite{Heil2008,Drotleff1993,Harissopulos2005}. The resulting fits with and without LUNA data are shown in Figure \ref{fig:sfactor}.  The reduced uncertainty on the \react{13}{C}{\an}{16}{O} cross section extrapolation at lower energies allows to reduce also the reaction rate error at stellar temperature of interest $T$~=~0.09 GK with respect to the NACRE II compilation\cite{Xu2013}. This allows to improve our understanding of the $s$-process branching points, in particular the ones sensitive to the neutron density. Inserting the new LUNA rate in sample AGB stellar models led to sizeable variations in the s-process yields of $^{60}$Fe, $^{152}$Gd and $^{205}$Pb~\cite{Ciani2021}. More extended AGB calculations still need to be performed to understand the possible influence of LUNA data on the galactic chemical evolution.  

The LUNA collaboration will extend the measurement of the \react{13}{C}{\an}{16}{O} to higher energies at the LUNA-MV facility~\cite{Ferraro2021}. This will give the unique opportunity of obtaining a complete dataset over a wide energy range, thus avoiding renormalization of other datasets with unknown systematic uncertainties.

\subsection{\react{22}{Ne}{\ag}{26}{Mg}} % Denise, Eliana
\begin{figure}[h!]
 \begin{center}
    \includegraphics[width=0.9\textwidth]{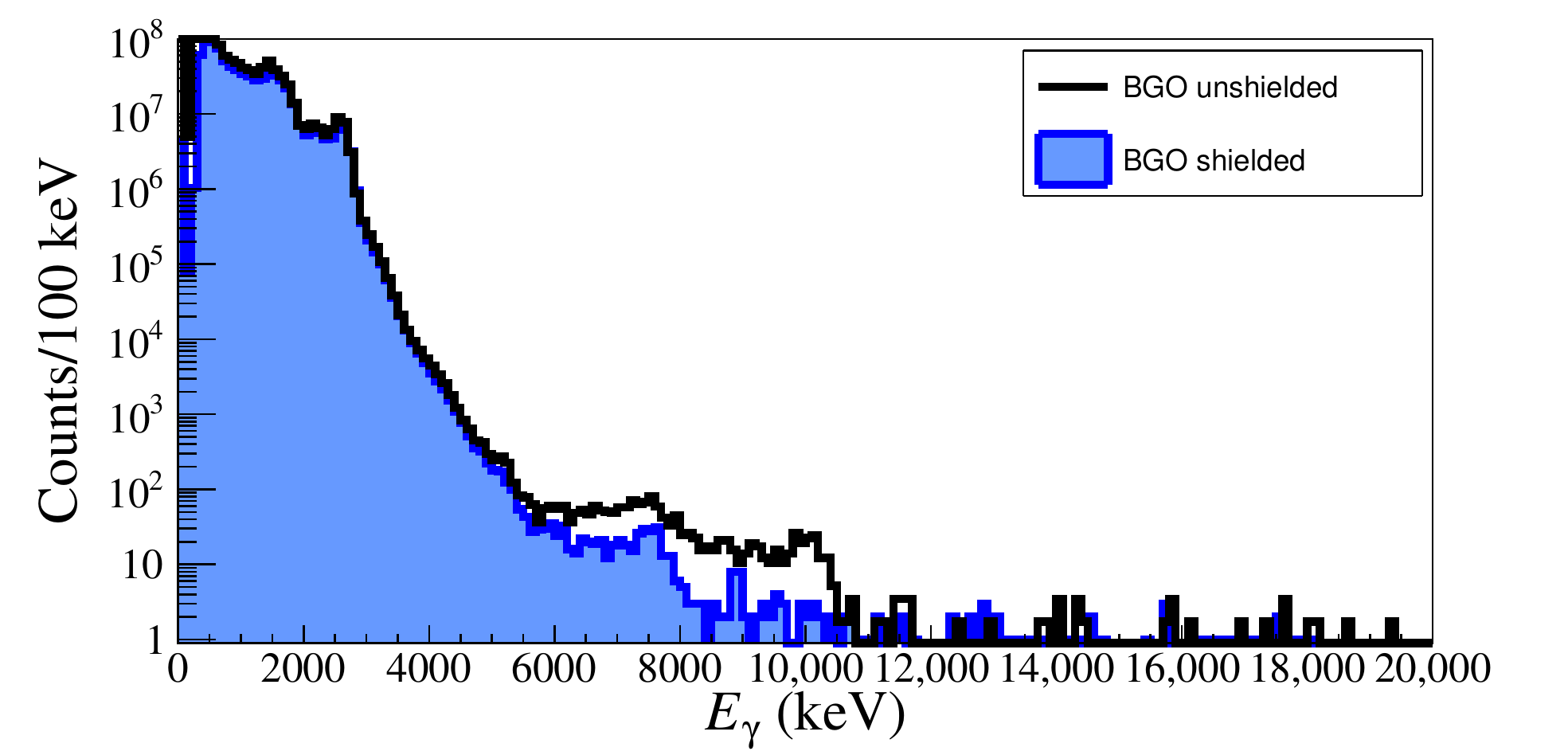}
    \caption{Background comparison between campaign I (black) and campaign II (blue).} %MDPI: For more than 4-digit numbers, comma should be added in the image, please revise -> DONE
    \label{fig1_22neag}
\end{center}
\end{figure}

$^{22}$Ne is produced during He-burning in massive~\cite{Thielemann-2018ASSL} and AGB stars~\cite{Lugaro-2018ASSL} stars via the sequence of reactions \react{14}{N}{\ag}{}{}$\react{18}{F}{(\beta^+,\nu)}{}{} \react{18}{O}{\ag}{22}{Ne}$ at typical temperatures of $\sim$$\SI{100}{MK}$. The $\alpha$-capture on \nuc{22}{Ne} proceeds through the reactions \react{22}{Ne}{\an}{25}{Mg} ($Q_{value} = \SI{-478}{keV}$) and 
\react{22}{Ne}{\ag}{26}{Mg} ($Q_{value} = \SI{10.6}{MeV}$).
In particular, the \react{22}{Ne}{\an}{25}{Mg} reaction is the main neutron source for the weak component of the $s$-process in massive stars. This process is responsible for the synthesis of nuclei between \nuc{56}{Fe} and \nuc{90}{Y}, and is presently less well understood compared with the main $s$-process. The weak $s$-process takes place at the end of the convective core He-burning, and at the beginning of carbon (shell) burning in massive stars where the temperatures reach \SIrange{0.22}{0.35}{GK}, high enough to activate the \react{22}{Ne}{\an}{25}{Mg} reaction. The contribution of the \react{22}{Ne}{\an}{25}{Mg} reaction to the neutron production is affected by the competitor \react{22}{Ne}{\ag}{26}{Mg} reaction. This last one, requiring lower temperatures (below \SI{300}{MK}), can be active during the entire He-burning phase, reducing the amount of the \nuc{22}{Ne} before the \react{22}{Ne}{\an}{25}{Mg} reaction is activated. Which of these reactions dominates depends on how their reaction rates evolve as a function of the temperature. Moreover, it has been observed that the \react{22}{Ne}{\ag}{26}{Mg} reaction rate affects also the nucleosynthesis of isotopes between \nuc{26}{Mg} and \nuc{31}{P} in intermediate-mass AGB stars~\cite{Karakas06-ApJ}. Therefore, to constrain the role of the \react{22}{Ne}{\an}{26}{Mg} reaction in the weak $s$-process the rate of both these reactions is required.
In the energy range of interest for different astrophysical scenarios, $\SI{250}{keV} \le E \le \SI{900}{keV}$, the \react{22}{Ne}{\ag}{26}{Mg} reaction rate is affected by several resonances, \Ecm\,=\,(334.4~$\pm$\,0.8)\,keV, (469\,$\pm$\,1)\,keV, (556.33\,$\pm$\,0.05)\,keV and (706\,$\pm$\,1)\,keV~\cite{Adsley2021}.
Except for the \Ecm\,=\,706\,keV resonance, which was measured directly for both channels~\cite{Jaeger01-PRL}, the other resonances have been studied only indirectly and the data on excitation energies and spin-parity assignments are still under debate.
Among these resonances the only one accessible at the LUNA \SI{400}{kV} accelerator is the $\Ecm$~=~\SI{334.4}{keV}, corresponding to $\Ex$ = {10,949}~{keV}  excited level of the \nuc{26}{Mg}. This resonance has a key role in determining the temperature at which the $\an$ starts to dominate. Indeed, the \SI{334.4}{keV} resonance is below the neutron threshold, $S_\mathrm{n}$, and thus contributes to the \react{22}{Ne}{\ag}{26}{Mg} reaction rate only. This resonance has been studied in several works~\cite{Giesen93-NPA, Ugalde07-PRC, Longland09-PRC, Talwar16-PRC, Lotay19-EPJA, Jayatissa-2020PhLB, Ota-2020PhLB}.
All these studies and subsequent evaluations~\cite{ANGULO1999, Iliadis10-NPA, Longland12-PRC, Sallaska13-ApJSS, Adsley2021} lead to a broad range of possible
values for the \SI{334}{keV} resonance strength ($10^{-14}$--$10^{-9}$\,eV), which significantly affects the \react{22}{Ne}{\ag}{26}{Mg} reaction rate.
At LUNA, the resonance was studied in two different campaigns using the intense $\alpha$-beam from the LUNA \SI{400}{kV} machine and the windowless gas target filled with  {99.99}\% enriched \nuc{22}{Ne} gas combined with a high efficiency BGO detector.
The first campaign was performed using the same experimental setup described in Section~\ref{22ne_pg_23na}~\cite{Ferraro2018-EPJ}. The data collected during the first campaign \cite{Piatti18-PhD} were affected by the environmental background. In particular, in the \Eg\,=\,\SIrange{6}{18}{MeV} region, the background deep underground mainly arises from neutron-induced reactions~\cite{Bemmerer-2005EPJA, Best16-EPJA, Boeltzig18-JPG}. In order to reduce this residual background, a \SI{10}{cm} thick shield, made of borated ({5}\%) polyethylene (PE-HMW 500 BOR5 by Profilan Kunstoffwerk), was added around the BGO. With this improvement, the background count rate in the region of interest for the \react{22}{Ne}{\ag}{26}{Mg} reaction was reduced by a factor of (3.4\,$\pm$\,0.3), see Figure~\ref{fig1_22neag}. Such background reduction can be considered as a milestone for deep underground direct measurements and provides a motivation for higher energy resonance studies that will be possible with LUNA-MV.

\subsection{\react{22}{Ne}{\an}{25}{Mg}} % Chemseddine, David

The \react{22}{Ne}{\an}{25}{Mg} reaction is the main source of neutrons for the weak s-process and also provides an important neutron burst during the main s-process. Its study is therefore crucial for the understanding of the stellar nucleosynthesis of some heavy elements in the Universe,
especially those with atomic mass $60<A<90$. To date, only upper limits are available for the \react{22}{Ne}{\an}{25}{Mg} cross section below the resonance at $\Ea = \SI{832}{keV}$ ($\Jpi = 2^+$)~\cite{Jaeger01-PRL,Harms91,Drotleff1993}. A number of natural parity states exist between the neutron threshold and the \SI{832}{keV} state, but their contribution to the cross section is unknown so far~\cite{Jaeger01-PRL,Bisterzo_2014,massimi12}.

\begin{figure}[h!]
    \centering
    \includegraphics[width=.85\textwidth]{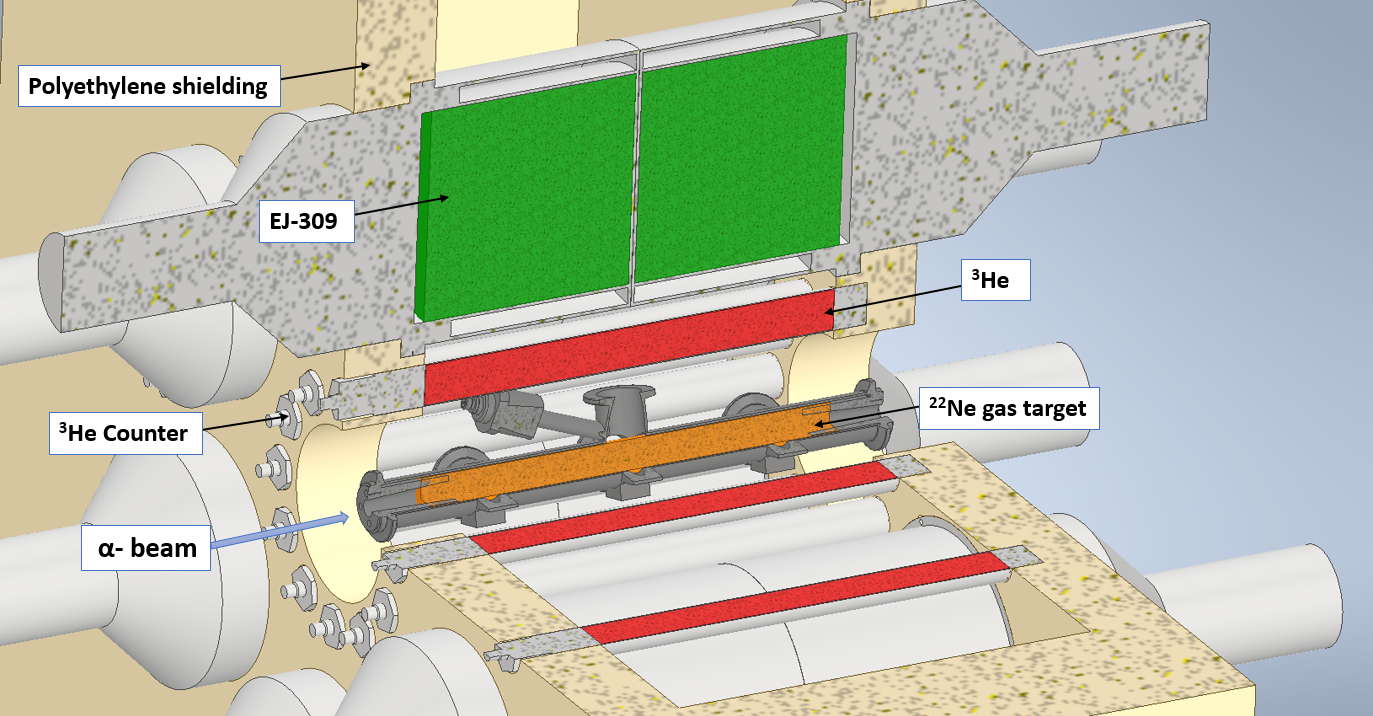}
    \caption{Cutoff view of the SHADES detector array surrounding the target chamber.}
    \label{fig:SHADES}
\end{figure}

As a first approach to overcome the challenging experimental conditions, the SHADES project (ERC-StG 2019 \#852016) aims to perform a low energy and high sensitivity measurement at the LNGS. Moving the measurement to a deep underground environment will increase the sensitivity by more than three orders of magnitude compared to the state of the art. Another novelty of this setup is the use of a hybrid detection system of 12 EJ-309 liquid scintillators and a total of 18 \nuc{3}{He} counters arranged in two rings surrounding a recirculating windowless gas target. The whole detection setup is located inside a shielding of borated polyethylene (Figure~\ref{fig:SHADES}). While the \nuc{3}{He} counters can only give the number of detected neutrons, the liquid scintillators thermalise the neutrons coming from the reaction occurring inside the gas target chamber, so they can be detected by the \nuc{3}{He} counters with higher efficiency. In addition, the scintillators can provide the energy loss of those same neutrons, allowing for the determination of their initial energies. 
This combination of multiple detector types will also unable a better discrimination against beam-induced background by imposing that only events detected in both detector types are considered. Furthermore, being able to get information about the neutron energy will allow to reject high energy neutrons coming from positive $Q$-value beam-induced reactions with elements such as carbon and boron. The \react{22}{Ne}{\an}{25}{Mg} reaction will be studied with the new LUNA-MV accelerator. The SHADES detector, combined with the high intensity He beam, will allow us to directly access the relevant astrophysical energies while aiming to reach the lowest accessible energies, well below the $\Ea~=~\SI{832}{keV}$ resonance state.

%%%%%%%%%%%%%%%%%%%%%%%%%%%%%%%%%%%%%%%%%%

\section{Conclusions} % Francesca
 
In the last 30 years the LUNA collaboration has made significant efforts to compose the mosaic made of hundreds of nuclear reactions that are crucial for stellar evolution and nucleosynthesis in AGB stars. In particular, LUNA has studied many reactions essential for the synthesis of the light elements in the hydrogen burning shell, as summarized in Table \ref{tab:reactions}. Direct data are now available for most of the reactions of the CNO cycle, and the measurements of proton capture cross sections on $^{17,18}$O improved the present knowledge on the expected oxygen isotopic ratios from AGB stars. In addition, the studies on the NeNa cycle contribute to a detailed understanding of the O-Na anti-correlation in Globular Clusters. More recently, LUNA has started the study of reactions involved in s-process nucleosynthesis. The new measurement of the \react{13}{C}{\an}{16}{O} cross section within the Gamow peak represents a milestone in the knowledge of neutrons sources. 

%%%%%%%%%%%%%%%%%%%%%%%%%%%%%%%%%%%%%%%%%%%%%%%%%%%%%%%%%%
     \begin{table}[h!]\footnotesize
 
        \caption{Overview of all reactions discussed in the present paper. The energy ranges covered at LUNA are in the centre-of-mass system and specific energy values refer to resonances. The table reports also an indication of the temperature ranges which are affected the most by the LUNA measurements.}
        \setlength{\tabcolsep}{8.1mm}
        \begin{tabular}{ c  c  c   c }
        \hline
         \textbf{Reaction} & \textbf{Burning Network} & {\bf LUNA Range} & \textbf{Relevant \emph{T}} \\
          & & \textbf{[keV]} & \textbf{[GK]} \\
        \hline
 
        $^{17}$O\pa{}$^{14}$N & CNO & 64.5, 183  & 0.05--0.24  \\
        $^{17}$O\pg{}$^{18}$F & CNO &  64.5, 167--370  &  0.05--0.68   \\
        $^{18}$O\pa{}$^{15}$N & CNO &  55--340  &  0.04--0.6  \\
        $^{18}$O\pg{}$^{19}$F & CNO &  85--150  &  0.07--0.18  \\
        $^{12}$C\pg{}$^{13}$N & CNO &  74--370  &  0.08--0.9  \\
        $^{13}$C\pg{}$^{14}$N & CNO &  74--370  &  0.08--0.9  \\
        $^{22}$Ne\pg{}$^{23}$Na & Ne-Na &  68--300  &  0.04--0.39  \\
        $^{20}$Ne\pg{}$^{21}$Na & Ne-Na &  366  & 0.53   \\
        $^{23}$Na\pg{}$^{24}$Mg & Ne-Na &  138, 240, 296  &   0.11--0.35  \\
        $^{13}$C($\alpha$,n)$^{16}$O & s-process &  230--300  &  0.12--0.18   \\ 
        $^{22}$Ne\ag{}$^{26}$Mg & s-process &  334  &   0.12  \\
         $^{22}$Ne\an{}$^{25}$Mg & s-process &   >350  &   >0.1   \\
        \hline
        \end{tabular}
    \label{tab:reactions}
   
    \end{table}
%%%%%%%%%%%%%%%%%%%%%%%%%%%%%%%%%%%%%%%%%%%%%%%%%%%%%%%%%%

Even if big steps forward have been made, there are still missing information and reactions requiring more accurate experimental data. For this reason, the new LUNA-MV facility at Gran Sasso represents a unique opportunity to study the nucleosynthesis in massive stars as well as their evolution and the production of about half of the chemical elements heavier than iron through the $s$-process.    

\vspace{6pt} 

%%%%%%%%%%%%%%%%%%%%%%%%%%%%%%%%%%%%%%%%%%

\paragraph{Funding}{This work was supported by INFN with contributions by other institutions as detailed in the acknowledgments section.}

\paragraph{Acknowledgments}Support from the National Research, Development and Innovation Office NKFIH, Hungary (contract number PD129060) is acknowledged. D.\,R. and C.\,A. acknowledge funding from the European Research Council (ERC-StG 2019 \#852016). Support from the Italian Ministry of Education, University and Research (MIUR) through the ``Dipartimenti di eccellenza'' project ``Science of the Universe'' and a DAAD fellowship at HZDR for D.\,P. are gratefully acknowledged. A.B. acknowledges ChETEC-INFRA (EU project \#101008324). G.\,F. C. acknowledges ECOST-STSM-Request-CA16117-43599. C.\,G. B. acknowledges support from UKRI STFC.

%%%%%%%%%%%%%%%%%%%%%%%%%%%%%%%%%%%%%%%%%%

% Please provide either the correct journal abbreviation (e.g., according to the “List of Title Word Abbreviations” http://www.issn.org/services/online-services/access-to-the-ltwa/) or the full name of the journal.


%\centering %% If there is a figure in wide page, please release command \centering
\bibliographystyle{unsrt}
\bibliography{AGB_review_LUNA}

\begin{thebibliography}{100}

\bibitem{Busso99-AnnRev}
M.~Busso, R.~Gallino, and G.~J. Wasserburg.
\newblock Nucleosynthesis in asymptotic giant branch stars: Relevance for
  galactic enrichment and solar system formation.
\newblock {\em Annual Review of Astronomy and Astrophysics}, 37(1):239--309,
  1999.

\bibitem{Greife1994}
U.~Greife, C.~Arpesella, C.A. Barnes, F.~Bartolucci, E.~Bellotti, C.~Broggini,
  P.~Corvisiero, G.~Fiorentini, A.~Fubini, G.~Gervino, F.~Gorris, C.~Gustavino,
  M.~Junker, R.W. Kavanagh, A.~Lanza, G.~Mezzorani, P.~Prati, P.~Quarati, W.S.
  Rodney, C.~Rolfs, W.H. Schulte, H.P. Trautvetter, and D.~Zahnow.
\newblock Laboratory for underground nuclear astrophysics (luna).
\newblock {\em Nucl. Instrum. Methods Phys. Res. A}, 350(1):327--337, 1994.

\bibitem{Formicola2003}
A.~Formicola, G.~Imbriani, M.~Junker, D.~Bemmerer, R.~Bonetti, C.~Broggini,
  C.~Casella, P.~Corvisiero, H.~Costantini, G.~Gervino, C.~Gustavino, A.~Lemut,
  P.~Prati, V.~Roca, C.~Rolfs, M.~Romano, D.~Sch\"urmann, F.~Strieder,
  F.~Terrasi, H.-P. Trautvetter, and S.~Zavatarelli.
\newblock The luna ii 400\,kv accelerator.
\newblock {\em Nucl. Instrum. Methods Phys. Res. A}, 507(3):609--616, 2003.

\bibitem{Cavanna2014}
F.~{Cavanna}, R.~{Depalo}, M.~L. {Menzel}, M.~{Aliotta}, M.~{Anders},
  D.~{Bemmerer}, C.~{Broggini}, C.~G. {Bruno}, A.~{Caciolli}, P.~{Corvisiero},
  T.~{Davinson}, A.~{di Leva}, Z.~{Elekes}, F.~{Ferraro}, A.~{Formicola}, Zs.
  {F{\"u}l{\"o}p}, G.~{Gervino}, A.~{Guglielmetti}, C.~{Gustavino}, Gy.
  {Gy{\"u}rky}, G.~{Imbriani}, M.~{Junker}, R.~{Menegazzo}, P.~{Prati},
  C.~{Rossi Alvarez}, D.~A. {Scott}, E.~{Somorjai}, O.~{Straniero},
  F.~{Strieder}, T.~{Sz{\"u}cs}, and D.~{Trezzi}.
\newblock {A new study of the
  ${}^{22}\mathrm{Ne}(\mathrm{p},\gamma){}^{23}\mathrm{Na}$ reaction deep
  underground: Feasibility, setup and first observation of the 186 keV
  resonance}.
\newblock {\em European Physical Journal A}, 50:179, 11 2014.

\bibitem{Ferraro2018-EPJ}
F.~{Ferraro}, M.~P. {Tak{\'a}cs}, D.~{Piatti}, V.~{Mossa}, M.~{Aliotta},
  D.~{Bemmerer}, A.~{Best}, A.~{Boeltzig}, C.~{Broggini}, C.~G. {Bruno},
  A.~{Caciolli}, F.~{Cavanna}, T.~{Chillery}, G.~F. {Ciani}, P.~{Corvisiero},
  L.~{Csedreki}, T.~{Davinson}, R.~{Depalo}, G.~{D'Erasmo}, A.~{Di Leva},
  Z.~{Elekes}, E.~M. {Fiore}, A.~{Formicola}, Zs. {F{\"u}l{\"o}p},
  G.~{Gervino}, A.~{Guglielmetti}, C.~{Gustavino}, Gy. {Gy{\"u}rky},
  G.~{Imbriani}, M.~{Junker}, I.~{Kochanek}, M.~{Lugaro}, L.~E. {Marcucci},
  P.~{Marigo}, R.~{Menegazzo}, F.~R. {Pantaleo}, V.~{Paticchio}, R.~{Perrino},
  P.~{Prati}, L.~{Schiavulli}, K.~{St{\"o}ckel}, O.~{Straniero},
  T.~{Sz{\"u}cs}, D.~{Trezzi}, and S.~{Zavatarelli}.
\newblock {A high-efficiency gas target setup for underground experiments, and
  redetermination of the branching ratio of the 189.5\,keV
  ${}^{22}\mathrm{Ne}(\mathrm{p},\gamma){}^{23}\mathrm{Na}$ resonance}.
\newblock {\em European Physical Journal A}, 54(3):44, 3 2018.

\bibitem{Mossa2020-EPJA}
V.~{Mossa}, K.~{St{\"o}ckel}, F.~{Cavanna}, F.~{Ferraro}, M.~{Aliotta},
  F.~{Barile}, D.~{Bemmerer}, A.~{Best}, A.~{Boeltzig}, C.~{Broggini}, C.~G.
  {Bruno}, A.~{Caciolli}, L.~{Csedreki}, T.~{Chillery}, G.~F. {Ciani},
  P.~{Corvisiero}, T.~{Davinson}, R.~{Depalo}, A.~{Di Leva}, Z.~{Elekes}, E.~M.
  {Fiore}, A.~{Formicola}, Zs. {F{\"u}l{\"o}p}, G.~{Gervino},
  A.~{Guglielmetti}, C.~{Gustavino}, G.~{Gy{\"u}rky}, G.~{Imbriani},
  M.~{Junker}, I.~{Kochanek}, M.~{Lugaro}, L.~E. {Marcucci}, P.~{Marigo},
  E.~{Masha}, R.~{Menegazzo}, F.~R. {Pantaleo}, V.~{Paticchio}, R.~{Perrino},
  D.~{Piatti}, P.~{Prati}, L.~{Schiavulli}, O.~{Straniero}, T.~{Sz{\"u}cs},
  M.~P. {Tak{\'a}cs}, D.~{Trezzi}, S.~{Zavatarelli}, and G.~{Zorzi}.
\newblock {Setup commissioning for an improved measurement of the
  $\mathrm{D}(\mathrm{p},\gamma){}^{3}\mathrm{He}$ cross section at Big Bang
  Nucleosynthesis energies}.
\newblock {\em European Physical Journal A}, 56(5):144, 5 2020.

\bibitem{Boeltzig18-JPG}
A.~{Boeltzig}, A.~{Best}, G.~{Imbriani}, M.~{Junker}, M.~{Aliotta},
  D.~{Bemmerer}, C.~{Broggini}, C.~G. {Bruno}, R.~{Buompane}, A.~{Caciolli},
  F.~{Cavanna}, T.~{Chillery}, G.~F. {Ciani}, P.~{Corvisiero}, L.~{Csedreki},
  T.~{Davinson}, R.~J. {deBoer}, R.~{Depalo}, A.~{Di Leva}, Z.~{Elekes},
  F.~{Ferraro}, E.~M. {Fiore}, A.~{Formicola}, Z.~{F{\"u}l{\"o}p},
  G.~{Gervino}, A.~{Guglielmetti}, C.~{Gustavino}, G.~{Gy{\"u}rky},
  I.~{Kochanek}, R.~{Menegazzo}, V.~{Mossa}, F.~R. {Pantaleo}, V.~{Paticchio},
  R.~{Perrino}, D.~{Piatti}, P.~{Prati}, L.~{Schiavulli}, K.~{St{\"o}ckel},
  O.~{Straniero}, F.~{Strieder}, T.~{Sz{\"u}cs}, M.~P. {Tak{\'a}cs},
  D.~{Trezzi}, M.~{Wiescher}, and S.~{Zavatarelli}.
\newblock Improved background suppression for radiative capture reactions at
  {LUNA} with {HPGe} and {BGO} detectors.
\newblock {\em J.\ Phys.\ G}, 45(2):025203, jan 2018.

\bibitem{Bruno2015}
C.~G. {Bruno}, D.~A. {Scott}, A.~{Formicola}, M.~{Aliotta}, T.~{Davinson},
  M.~{Anders}, A.~{Best}, D.~{Bemmerer}, C.~{Broggini}, A.~{Caciolli},
  F.~{Cavanna}, P.~{Corvisiero}, R.~{Depalo}, A.~{Di Leva}, Z.~{Elekes}, Zs.
  {F{\"u}l{\"o}p}, G.~{Gervino}, C.~J. {Griffin}, A.~{Guglielmetti},
  C.~{Gustavino}, Gy. {Gy{\"u}rky}, G.~{Imbriani}, M.~{Junker}, R.~{Menegazzo},
  E.~{Napolitani}, P.~{Prati}, E.~{Somorjai}, O.~{Straniero}, F.~{Strieder},
  T.~{Sz{\"u}cs}, and D.~{Trezzi}.
\newblock {Resonance strengths in the
  ${}^{17,18}\mathrm{O}(\mathrm{p},\alpha){}^{14,15}\mathrm{N}$ reactions and
  background suppression underground. Commissioning of a new setup for
  charged-particle detection at LUNA}.
\newblock {\em European Physical Journal A}, 51:94, 8 2015.

\bibitem{Balibrea2018}
J.~Balibrea-Correa, G.F. Ciani, R.~Buompane, F.~Cavanna, L.~Csedreki,
  R.~Depalo, F.~Ferraro, and A.~Best.
\newblock {Improved pulse shape discrimination for high pressure
  ${}^3\mathrm{He}$ counters}.
\newblock {\em Nucl. Instrum. Methods Phys. Res. A}, 906:103, 2018.

\bibitem{Csedreki2021}
L.~{Csedreki}, G.~F. {Ciani}, J.~{Balibrea-Correa}, A.~{Best}, M.~{Aliotta},
  F.~{Barile}, D.~{Bemmerer}, A.~{Boeltzig}, C.~{Broggini}, C.~G. {Bruno},
  A.~{Caciolli}, F.~{Cavanna}, T.~{Chillery}, P.~{Colombetti}, P.~{Corvisiero},
  T.~{Davinson}, R.~{Depalo}, A.~{Di Leva}, Z.~{Elekes}, F.~{Ferraro}, E.~M.
  {Fiore}, A.~{Formicola}, Zs. {F{\"u}l{\"o}p}, G.~{Gervino},
  A.~{Guglielmetti}, C.~{Gustavino}, Gy. {Gy{\"u}rky}, G.~{Imbriani},
  Z.~{Janas}, M.~{Junker}, I.~{Kochanek}, M.~{Lugaro}, P.~{Marigo}, E.~{Masha},
  C.~{Mazzocchi}, R.~{Menegazzo}, V.~{Mossa}, F.~R. {Pantaleo}, V.~{Paticchio},
  R.~{Perrino}, D.~{Piatti}, P.~{Prati}, L.~{Schiavulli}, K.~{St{\"o}ckel},
  O.~{Straniero}, T.~{Sz{\"u}cs}, M.~P. {Tak{\'a}cs}, F.~{Terrasi}, and
  S.~{Zavatarelli}.
\newblock {Characterization of the LUNA neutron detector array for the
  measurement of the ${}^{13}\mathrm{C}(\alpha, \mathrm{n}){}^{16}\mathrm{O}$
  reaction}.
\newblock {\em Nucl. Instrum. Methods Phys. Res. A}, 994:165081, 4 2021.

\bibitem{Cavanna18-IJMPA}
F.~{Cavanna} and P.~{Prati}.
\newblock {Direct measurement of nuclear cross-section of astrophysical
  interest: Results and perspectives}.
\newblock {\em Int. J. Mod. Phys. A}, 33(9):1843010--346, Mar 2018.

\bibitem{Broggini2019}
C.~{Broggini}, O.~{Straniero}, M.~G.~F. {Taiuti}, G.~{de Angelis},
  G.~{Benzoni}, G.~E. {Bruno}, S.~{Bufalino}, G.~{Cardella}, N.~{Colonna},
  M.~{Contalbrigo}, G.~{Cosentino}, S.~{Cristallo}, C.~{Curceanu}, E.~{De
  Filippo}, R.~{Depalo}, A.~{Di Leva}, A.~{Feliciello}, S.~{Gammino},
  A.~{Galat{\`a}}, M.~{La Cognata}, R.~{Lea}, S.~{Leoni}, I.~{Lombardo},
  V.~{Manzari}, D.~{Mascali}, C.~{Massimi}, A.~{Mengoni}, D.~{Mengoni}, D.~R.
  {Napoli}, S.~{Palmerini}, S.~{Piano}, S.~{Pirrone}, R.~G. {Pizzone},
  G.~{Politi}, P.~{Prati}, G.~{Prete}, P.~{Russotto}, G.~{Tagliente}, and G.~M.
  {Urciuoli}.
\newblock {Experimental nuclear astrophysics in Italy}.
\newblock {\em La Rivista del Nuovo Cimento}, 42:103, 3 2019.

\bibitem{Ferraro2021}
F.~{Ferraro}, G.~F. {Ciani}, A.~{Boeltzig}, F.~{Cavanna}, and S.~{Zavatarelli}.
\newblock {The study of key reactions shaping the post main-sequence evolution
  of massive stars in underground facilities}.
\newblock {\em Frontiers in Astronomy and Space Sciences}, 7:119, 3 2021.

\bibitem{Sen2019}
A.~{Sen}, G.~{Dom{\'\i}nguez-Ca{\~n}izares}, N.~C. {Podaru}, D.~J.~W. {Mous},
  M.~{Junker}, G.~{Imbriani}, and V.~{Rigato}.
\newblock {A high intensity, high stability 3.5\,MV Singletron{\texttrademark}
  accelerator}.
\newblock {\em Nucl. Instrum. Methods Phys. Res. B}, 450:390--395, 7 2019.

\bibitem{B2FH}
E.~M. {Burbidge}, G.~R. {Burbidge}, W.~A. {Fowler}, and F.~{Hoyle}.
\newblock {Synthesis} of the {Elements} in {Stars}.
\newblock {\em Rev.\ Mod.\ Phys.}, 29(4):547--650, 1 1957.

\bibitem{Berheide92}
M.~{Berheide}, C.~{Rolfs}, U.~{Schr{\"o}der}, and H.~P. {Trautvetter}.
\newblock {Search for the 70\,keV resonance in
  ${}^{17}\mathrm{O}(\mathrm{p},\alpha){}^{14}\mathrm{N}$}.
\newblock {\em Zeitschrift f{\"u}r Physik A Hadrons and nuclei},
  343(4):483--487, 1992.

\bibitem{Niemeyer96}
S.~{Niemeyer}.
\newblock Untersuchungen zur linienform der alpha-teilchen bei der
  ${}^{17}\mathrm{O}(\mathrm{p},\alpha){}^{14}\mathrm{N}$-reaktion.
\newblock Diplomarbeit, Ruhr-Universit\"at Bochum, 1996.

\bibitem{Blackmon95}
J.~C. {Blackmon}, A.~E. {Champagne}, M.~A. {Hofstee}, M.~S. {Smith}, R.~G.
  {Downing}, and G.~P. {Lamaze}.
\newblock {Measurement of the
  ${}^{17}\mathrm{O}(\mathrm{p},\alpha){}^{14}\mathrm{N}$ Cross Section at
  Stellar Energies}.
\newblock {\em Phys.\ Rev.\ Lett.}, 74(14):2642--2645, 4 1995.

\bibitem{Hannam99}
Mark~D. Hannam and William~J. Thompson.
\newblock Estimating small signals by using maximum likelihood and poisson
  statistics.
\newblock {\em Nucl. Instrum. Methods Phys. Res. A}, 431(1):239--251, 1999.

\bibitem{Sergi15}
M.~L. {Sergi}, C.~{Spitaleri}, M.~{La Cognata}, L.~{Lamia}, R.~G. {Pizzone},
  G.~G. {Rapisarda}, A.~{Mukhamedzhanov}, B.~{Irgaziev}, X.~D. {Tang},
  M.~{Wiescher}, J.~{Mrazek}, and V.~{Kroha}.
\newblock {Resonance strength measurement at astrophysical energies: The
  ${}^{17}\mathrm{O}(\mathrm{p},\alpha){}^{14}\mathrm{N}$ reaction studied via
  Trojan Horse Method}.
\newblock In {\em Nuclear Structure and Dynamics '15}, volume 1681 of {\em
  American Institute of Physics Conference Series}, page 050005, 2015.

\bibitem{Caciolli2012}
A.~{Caciolli}, D.~A. {Scott}, A.~{Di Leva}, A.~{Formicola}, M.~{Aliotta},
  M.~{Anders}, A.~{Bellini}, D.~{Bemmerer}, C.~{Broggini}, M.~{Campeggio},
  P.~{Corvisiero}, R.~{Depalo}, Z.~{Elekes}, Zs. {F{\"u}l{\"o}p}, G.~{Gervino},
  A.~{Guglielmetti}, C.~{Gustavino}, Gy. {Gy{\"u}rky}, G.~{Imbriani},
  M.~{Junker}, M.~{Marta}, R.~{Menegazzo}, E.~{Napolitani}, P.~{Prati},
  V.~{Rigato}, V.~{Roca}, C.~{Rolfs}, C.~{Rossi Alvarez}, E.~{Somorjai},
  C.~{Salvo}, O.~{Straniero}, F.~{Strieder}, T.~{Sz{\"u}cs}, F.~{Terrasi},
  H.~P. {Trautvetter}, and D.~{Trezzi}.
\newblock {Preparation and characterisation of isotopically enriched
  Ta$_{2}$O$_{5}$ targets for nuclear astrophysics studies}.
\newblock {\em European Physical Journal A}, 48:144, 10 2012.

\bibitem{Bruno2016}
C.~G. {Bruno}, D.~A. {Scott}, M.~{Aliotta}, A.~{Formicola}, A.~{Best},
  A.~{Boeltzig}, D.~{Bemmerer}, C.~{Broggini}, A.~{Caciolli}, F.~{Cavanna},
  G.~F. {Ciani}, P.~{Corvisiero}, T.~{Davinson}, R.~{Depalo}, A.~{Di Leva},
  Z.~{Elekes}, F.~{Ferraro}, Zs. {F{\"u}l{\"o}p}, G.~{Gervino},
  A.~{Guglielmetti}, C.~{Gustavino}, Gy. {Gy{\"u}rky}, G.~{Imbriani},
  M.~{Junker}, R.~{Menegazzo}, V.~{Mossa}, F.~R. {Pantaleo}, D.~{Piatti},
  P.~{Prati}, E.~{Somorjai}, O.~{Straniero}, F.~{Strieder}, T.~{Sz{\"u}cs},
  M.~P. {Tak{\'a}cs}, D.~{Trezzi}, and {LUNA Collaboration}.
\newblock {Improved Direct Measurement of the 64.5\,keV Resonance Strength in
  the ${}^{17}\mathrm{O}(\mathrm{p},\alpha){}^{14}\mathrm{N}$ Reaction at
  LUNA}.
\newblock {\em Phys.\ Rev.\ Lett.}, 117(14):142502, 9 2016.

\bibitem{Straniero2017}
O.~{Straniero}, C.~G. {Bruno}, M.~{Aliotta}, A.~{Best}, A.~{Boeltzig},
  D.~{Bemmerer}, C.~{Broggini}, A.~{Caciolli}, F.~{Cavanna}, G.~F. {Ciani},
  P.~{Corvisiero}, S.~{Cristallo}, T.~{Davinson}, R.~{Depalo}, A.~{Di Leva},
  Z.~{Elekes}, F.~{Ferraro}, A.~{Formicola}, Zs. {F{\"u}l{\"o}p}, G.~{Gervino},
  A.~{Guglielmetti}, C.~{Gustavino}, G.~{Gy{\"u}rky}, G.~{Imbriani},
  M.~{Junker}, R.~{Menegazzo}, V.~{Mossa}, F.~R. {Pantaleo}, D.~{Piatti},
  L.~{Piersanti}, P.~{Prati}, E.~{Samorjai}, F.~{Strieder}, T.~{Sz{\"u}cs},
  M.~P. {Tak{\'a}cs}, and D.~{Trezzi}.
\newblock {The impact of the revised
  ${}^{17}\mathrm{O}(\mathrm{p},\alpha){}^{14}\mathrm{N}$ reaction rate on
  ${}^{17}\mathrm{O}$ stellar abundances and yields}.
\newblock {\em Astronomy and Astrophysics}, 598:A128, 2 2017.

\bibitem{Lugaro2017}
M.~{Lugaro}, A.~I. {Karakas}, C.~G. {Bruno}, M.~{Aliotta}, L.~R. {Nittler},
  D.~{Bemmerer}, A.~{Best}, A.~{Boeltzig}, C.~{Broggini}, A.~{Caciolli},
  F.~{Cavanna}, G.~F. {Ciani}, P.~{Corvisiero}, T.~{Davinson}, R.~{Depalo},
  A.~{di Leva}, Z.~{Elekes}, F.~{Ferraro}, A.~{Formicola}, Zs. {F{\"u}l{\"o}p},
  G.~{Gervino}, A.~{Guglielmetti}, C.~{Gustavino}, Gy. {Gy{\"u}rky},
  G.~{Imbriani}, M.~{Junker}, R.~{Menegazzo}, V.~{Mossa}, F.~R. {Pantaleo},
  D.~{Piatti}, P.~{Prati}, D.~A. {Scott}, O.~{Straniero}, F.~{Strieder},
  T.~{Sz{\"u}cs}, M.~P. {Tak{\'a}cs}, and D.~{Trezzi}.
\newblock {Origin of meteoritic stardust unveiled by a revised proton-capture
  rate of ${}^{17}\mathrm{O}$}.
\newblock {\em Nature Astron.}, 1:0027, 1 2017.

\bibitem{Palmerini2021}
Sara Palmerini, Sergio Cristallo, Luciano Piersanti, Diego Vescovi, and
  Maurizio Busso.
\newblock Group ii oxide grains: How massive are their agb star progenitors?
\newblock {\em Universe}, 7(6), 2021.

\bibitem{Busso2010}
M.~{Busso}, S.~{Palmerini}, E.~{Maiorca}, S.~{Cristallo}, O.~{Straniero},
  C.~{Abia}, R.~{Gallino}, and M.~{La Cognata}.
\newblock {On the Need for Deep-mixing in Asymptotic Giant Branch Stars of Low
  Mass}.
\newblock {\em Astrophys.\ J.\ Lett.}, 717(1):L47--L51, 7 2010.

\bibitem{Scott2012}
D.~A. Scott, A.~Caciolli, A.~Di~Leva, A.~Formicola, M.~Aliotta, M.~Anders,
  D.~Bemmerer, C.~Broggini, M.~Campeggio, P.~Corvisiero, Z.~Elekes, Zs.
  F\"ul\"op, G.~Gervino, A.~Guglielmetti, C.~Gustavino, Gy. Gy\"urky,
  G.~Imbriani, M.~Junker, M.~Laubenstein, R.~Menegazzo, M.~Marta,
  E.~Napolitani, P.~Prati, V.~Rigato, V.~Roca, E.~Somorjai, C.~Salvo,
  O.~Straniero, F.~Strieder, T.~Sz\"ucs, F.~Terrasi, and D.~Trezzi.
\newblock First direct measurement of the
  ${}^{17}\mathrm{O}(\mathrm{p},\gamma)^{18}\mathrm{F}$ reaction cross section
  at gamow energies for classical novae.
\newblock {\em Phys.\ Rev.\ Lett.}, 109:202501, 11 2012.

\bibitem{DiLeva2014}
A.~Di~Leva, D.~A. Scott, A.~Caciolli, A.~Formicola, F.~Strieder, M.~Aliotta,
  M.~Anders, D.~Bemmerer, C.~Broggini, P.~Corvisiero, Z.~Elekes, Zs. F\"ul\"op,
  G.~Gervino, A.~Guglielmetti, C.~Gustavino, Gy. Gy\"urky, G.~Imbriani,
  J.~Jos\'e, M.~Junker, M.~Laubenstein, R.~Menegazzo, E.~Napolitani, P.~Prati,
  V.~Rigato, V.~Roca, E.~Somorjai, C.~Salvo, O.~Straniero, T.~Sz\"ucs,
  F.~Terrasi, and D.~Trezzi.
\newblock Underground study of the
  ${}^{17}\mathrm{O}(\mathrm{p},\gamma){}^{18}\mathrm{F}$ reaction relevant for
  explosive hydrogen burning.
\newblock {\em Phys.\ Rev.\ C}, 89:015803, 1 2014.

\bibitem{Laubenstein04}
M~Laubenstein, M~Hult, J~Gasparro, D~Arnold, S~Neumaier, G~Heusser, M~K\"ohler,
  P~Povinec, J.-L Reyss, M~Schwaiger, and P~Theodórsson.
\newblock Underground measurements of radioactivity.
\newblock {\em Applied Radiation and Isotopes}, 61(2):167--172, 2004.
\newblock Low Level Radionuclide Measurement Techniques - ICRM.

\bibitem{Tsuji-2008}
{Tsuji, T.}
\newblock Cool luminous stars: the hybrid nature of their infrared spectra.
\newblock {\em Astronomy and Astrophysics}, 489(3):1271--1289, 2008.

\bibitem{Meyer-2008}
Bradley~S. Meyer, Larry~R. Nittler, Ann~N. Nguyen, and Scott Messenger.
\newblock {Nucleosynthesis and Chemical Evolution of Oxygen}.
\newblock {\em Reviews in Mineralogy and Geochemistry}, 68(1):31--53, 01 2008.

\bibitem{Mak80}
H.~B. {Mak}, G.~T. {Ewan}, H.~C. {Evans}, J.~D. {MacArthur}, W.~{McLatchie},
  and R.~E. {Azuma}.
\newblock {The alpha widths of the 5603, 5605 and 5668\,keV states in
  ${}^{18}\mathrm{F}$}.
\newblock {\em Nucl. Phys. A}, 343:79--90, 7 1980.

\bibitem{Landre89}
V.~{Landre}, P.~{Aguer}, G.~{Bogaert}, A.~{Lefebvre}, J.~P. {Thibaud},
  S.~{Fortier}, J.~M. {Maison}, and J.~{Vernotte}.
\newblock {${}^{17}\mathrm{O}{}^{3}\mathrm{He},\mathrm{d}){}^{18}\mathrm{F}$
  reaction and its implication in the ${}^{17}\mathrm{O}$ destruction in the
  CNO cycle in stars}.
\newblock {\em Phys.\ Rev.\ C}, 40(5):1972--1984, 11 1989.

\bibitem{Fox2005}
C.~{Fox}, C.~{Iliadis}, A.~E. {Champagne}, R.~P. {Fitzgerald}, R.~{Longland},
  J.~{Newton}, J.~{Pollanen}, and R.~{Runkle}.
\newblock {Thermonuclear reaction rate of
  ${}^{17}\mathrm{O}(\mathrm{p},\gamma)^{18}\mathrm{F}$}.
\newblock {\em Phys.\ Rev.\ C}, 71(5):055801, 5 2005.

\bibitem{Gilmore08-Book}
G.~Gilmore.
\newblock {\em {{Practical $\gamma$-ray spectrometry, 2nd edition}}}.
\newblock John Wiley and Sons, New York, 2008.

\bibitem{Lacognata10}
M.~La Cognata, C.~Spitaleri, and A.~M. Mukhamedzhanov.
\newblock Effect of high-energy resonances on the
  ${}^{18}\mathrm{O}(\mathrm{p},\alpha){}^{15}\mathrm{N}$ reaction rate at agb
  and post-agb relevant temperatures.
\newblock {\em Astrophys.\ J.}, 723(2):1512--1522, oct 2010.

\bibitem{Mak78}
H.~B. {Mak}, H.~C. {Evans}, G.~T. {Ewan}, and J.~D. {Macarthur}.
\newblock {The ${}^{18}\mathrm{O}(\mathrm{p},\alpha){}^{15}\mathrm{N}$ cross
  section at low energies}.
\newblock {\em Nucl. Phys. A}, 304(1):210--220, 1978.

\bibitem{LW79}
H.~{Lorenz-Wirzba}, P.~{Schmalbrock}, H.~P. {Trautvetter}, M.~{Wiescher},
  C.~{Rolfs}, and W.~S. {Rodney}.
\newblock {The ${}^{18}\mathrm{O}(\mathrm{p}, \alpha){}^{15}\mathrm{N}$
  reaction at stellar energies}.
\newblock {\em Nucl. Phys. A}, 313(3):346--362, 1979.

\bibitem{Bruno2019}
C.G. Bruno, M.~Aliotta, P.~Descouvemont, A.~Best, T.~Davinson, D.~Bemmerer,
  A.~Boeltzig, C.~Broggini, A.~Caciolli, F.~Cavanna, T.~Chillery, G.F. Ciani,
  P.~Corvisiero, R.~Depalo, A.~{Di Leva}, Z.~Elekes, F.~Ferraro, A.~Formicola,
  Zs. F\"ul\"op, G.~Gervino, A.~Guglielmetti, C.~Gustavino, Gy. Gy\"urky,
  G.~Imbriani, M.~Junker, M.~Lugaro, P.~Marigo, R.~Menegazzo, V.~Mossa, F.R.
  Pantaleo, D.~Piatti, P.~Prati, K.~St\"ockel, O.~Straniero, F.~Strieder,
  T.~Sz\"ucs, M.P. Tak\'acs, and D.~Trezzi.
\newblock Improved astrophysical rate for the
  ${}^{18}\mathrm{O}(\mathrm{p},\alpha){}^{15}\mathrm{N}$ reaction by
  underground measurements.
\newblock {\em Phys.\ Lett.\ B}, 790:237--242, 2019.

\bibitem{Buckner2012}
M.~Q. Buckner, C.~Iliadis, J.~M. Cesaratto, C.~Howard, T.~B. Clegg, A.~E.
  Champagne, and S.~Daigle.
\newblock Thermonuclear reaction rate of
  ${}^{18}\mathrm{O}(\mathrm{p},\gamma){}^{19}\mathrm{F}$.
\newblock {\em Phys.\ Rev.\ C}, 86(6):065804, 12 2012.

\bibitem{Fortune2013}
H.~T. Fortune.
\newblock Resonance-strength parameter for
  ${}^{18}\mathrm{O}(\mathrm{p},\gamma)$ at $e_\mathrm{p} = 90\,\mathrm{keV}$.
\newblock {\em Phys.\ Rev.\ C}, 88(1):015801, 7 2013.

\bibitem{Best2019}
A.~Best, F.R. Pantaleo, A.~Boeltzig, G.~Imbriani, M.~Aliotta,
  J.~Balibrea-Correa, D.~Bemmerer, C.~Broggini, C.G. Bruno, R.~Buompane,
  A.~Caciolli, F.~Cavanna, T.~Chillery, G.F. Ciani, P.~Corvisiero, L.~Csedreki,
  T.~Davinson, R.J. deBoer, R.~Depalo, A.~{Di Leva}, Z.~Elekes, F.~Ferraro,
  E.M. Fiore, A.~Formicola, Zs. F\"ul\"op, G.~Gervino, A.~Guglielmetti,
  C.~Gustavino, Gy. Gy\"urky, M.~Junker, I.~Kochanek, M.~Lugaro, P.~Marigo,
  R.~Menegazzo, V.~Mossa, V.~Paticchio, R.~Perrino, D.~Piatti, P.~Prati,
  L.~Schiavulli, K.~St\"ockel, O.~Straniero, F.~Strieder, T.~Sz\"ucs, M.P.
  Tak\'acs, D.~Trezzi, M.~Wiescher, and S.~Zavatarelli.
\newblock Cross section of the reaction
  ${}^{18}\mathrm{O}(\mathrm{p},\gamma)^{19}\mathrm{F}$ at astrophysical
  energies: The 90\,kev resonance and the direct capture component.
\newblock {\em Phys.\ Lett.\ B}, 797:134900, 2019.

\bibitem{Wiescher1980}
M.~Wiescher, H.W. Becker, J.~G\"orres, K.-U. Kettner, H.P. Trautvetter, W.E.
  Kieser, C.~Rolfs, R.E. Azuma, K.P. Jackson, and J.W. Hammer.
\newblock Nuclear and astrophysical aspects of
  ${}^{18}\mathrm{O}(\mathrm{p},\gamma){}^{19}\mathrm{F}$.
\newblock {\em Nucl. Phys. A}, 349(1-2):165--216, 11 1980.

\bibitem{Pantaleo2021}
F.~R. Pantaleo, A.~Boeltzig, A.~Best, R.~Perrino, M.~Aliotta,
  J.~Balibrea-Correa, F.~Barile, D.~Bemmerer, C.~Broggini, C.~G. Bruno,
  R.~Buompane, A.~Caciolli, F.~Cavanna, T.~Chillery, G.~F. Ciani,
  P.~Corvisiero, L.~Csedreki, T.~Davinson, R.~J. deBoer, R.~Depalo,
  G.~D'Erasmo, A.~Di~Leva, Z.~Elekes, F.~Ferraro, E.~M. Fiore, A.~Formicola,
  Zs. F\"ul\"op, G.~Gervino, A.~Guglielmetti, C.~Gustavino, Gy. Gy\"urky,
  G.~Imbriani, M.~Junker, I.~Kochanek, M.~Lugaro, E.~Masha, R.~Menegazzo,
  V.~Mossa, V.~Paticchio, D.~Piatti, P.~Prati, D.~Rapagnani, L.~Schiavulli,
  K.~St\"ockel, O.~Straniero, T.~Sz\"ucs, M.~P. Tak\'acs, D.~Trezzi,
  M.~Wiescher, and S.~Zavatarelli.
\newblock Low-energy resonances in the
  ${}^{18}\mathrm{O}(\mathrm{p},\gamma){}^{19}\mathrm{F}$ reaction.
\newblock {\em Phys.\ Rev.\ C}, 104(2):025802, 8 2021.

\bibitem{Savage2008}
Chandra Savage, A.~Apponi, L.~Ziurys, and and Wyckoff.
\newblock Galactic ${}^{12}\mathrm{C}/{}^{13}\mathrm{C}$ ratios from
  millimeter-wave observations of interstellar cn.
\newblock {\em Astrophys.\ J.}, 578:211, 12 2008.

\bibitem{Palmerini2011}
S.~{Palmerini}, M.~{La Cognata}, S.~{Cristallo}, and M.~{Busso}.
\newblock {Deep Mixing in Evolved Stars. I. The Effect of Reaction Rate
  Revisions from C to Al}.
\newblock {\em Astrophys.\ J.}, 729(1):3, 3 2011.

\bibitem{Liu2021}
Nan {Liu}, Jens {Barosch}, Larry~R. {Nittler}, Conel~M. {O'D. Alexander},
  Jianhua {Wang}, Sergio {Cristallo}, Maurizio {Busso}, and Sara {Palmerini}.
\newblock {New Multielement Isotopic Compositions of Presolar SiC Grains:
  Implications for Their Stellar Origins}.
\newblock {\em Astrophys.\ J.\ Lett.}, 920(1):L26, 10 2021.

\bibitem{Palmerini2017}
S.~{Palmerini}, O.~{Trippella}, and M.~{Busso}.
\newblock {A deep mixing solution to the aluminum and oxygen isotope puzzles in
  pre-solar grains}.
\newblock {\em Monthly Notices of the Royal Astronomical Society},
  467(1):1193--1201, 5 2017.

\bibitem{Baily1950}
Carl~L. Bailey and William~R. Stratton.
\newblock {Cross Section of the
  $\mathrm{C}^{12}(\mathrm{p},\gamma)\mathrm{N}^{13}$ Reaction at Low
  Energies}.
\newblock {\em Phys. Rev.}, 77:194--196, Jan 1950.

\bibitem{Lamb1957}
W.~A.~S. Lamb and Ross~E. Hester.
\newblock {Radiative Capture of Protons in Carbon from 80 to 126\,kev}.
\newblock {\em Phys. Rev.}, 107:550--553, Jul 1957.

\bibitem{Rolfs1974}
C.~Rolfs and R.E. Azuma.
\newblock {Interference effects in
  ${}^{12}\mathrm{C}(\mathrm{p},\gamma){}^{13}\mathrm{N}$ and direct capture to
  unbound states}.
\newblock {\em Nucl. Phys. A}, 227(2):291 -- 308, 1974.

\bibitem{Burtebaev2008}
N.~Burtebaev, S.~B. Igamov, R.~J. Peterson, R.~Yarmukhamedov, and D.~M.
  Zazulin.
\newblock New measurements of the astrophysical $s$-factor for
  ${}^{12}\mathrm{C}(\mathrm{p},\gamma){}^{13}\mathrm{N}$ reaction at low
  energies and the asymptotic normalization coefficient (nuclear vertex
  constant) for the $\mathrm{p} + {}^{12}\mathrm{C} \to {}^{13}\mathrm{N}$
  reaction.
\newblock {\em Phys.\ Rev.\ C}, 78(3):035802, 9 2008.

\bibitem{Vogl1963}
J.~L. Vogl.
\newblock {\em Radiative capture of protons by ${}^{12}\mathrm{C}$ and
  ${}^{13}\mathrm{C}$ below 700\,keV}.
\newblock PhD thesis, California Institute of Technology, 1 1963.

\bibitem{King1994}
J.~D. {King}, R.~E. {Azuma}, J.~B. {Vise}, J.~{G{\"o}rres}, C.~{Rolfs}, H.~P.
  {Trautvetter}, and A.~E. {Vlieks}.
\newblock Cross section and astrophysical s-factor for the
  ${}^{13}\mathrm{C}(\mathrm{p}, \gamma){}^{14}\mathrm{N}$ reaction.
\newblock {\em Nucl. Phys. A}, 567(2):354--376, Jan 1994.

\bibitem{Seagrave1952}
John~D. {Seagrave}.
\newblock {Radiative Capture of Protons by $\mathrm{C}{13}$}.
\newblock {\em Phys. Rev.}, 85(2):197--203, 1 1952.

\bibitem{Woodbury1952}
Eric~John {Woodbury} and William~Alfred {Fowler}.
\newblock {The Cross Section for the Radiative Capture of Protons by
  $\mathrm{C}^{13}$ at 129\,kev}.
\newblock {\em Phys. Rev.}, 85(1):51--57, 1 1952.

\bibitem{Hester1961}
Ross~E. {Hester} and W.~A. {Lamb}.
\newblock {Radiative Capture of Protons in $\mathrm{C}^{13}$}.
\newblock {\em Phys. Rev.}, 121(2):584--586, 1 1961.

\bibitem{Genard2010}
G~Genard, P~Descouvemont, and G~Terwagne.
\newblock S-factor measurement of the
  ${}^{13}\mathrm{C}(\mathrm{p},\gamma){}^{14}\mathrm{N}$ reaction in reverse
  kinematics.
\newblock {\em J.\ Phys. Conf.\ Ser.}, 202:012015, 1 2010.

\bibitem{Ciani2020}
G.~F. Ciani, L.~Csedreki, J.~Balibrea-Correa, A.~Best, M.~Aliotta, F.~Barile,
  D.~Bemmerer, A.~Boeltzig, C.~Broggini, C.~G. Bruno, A.~Caciolli, F.~Cavanna,
  T.~Chillery, P.~Colombetti, P.~Corvisiero, T.~Davinson, R.~Depalo,
  A.~Di~Leva, L.~Di~Paolo, Z.~Elekes, F.~Ferraro, E.~M. Fiore, A.~Formicola,
  Zs. F{\"u}l{\"o}p, G.~Gervino, A.~Guglielmetti, C.~Gustavino, Gy. Gy{\"u}rky,
  G.~Imbriani, M.~Junker, I.~Kochanek, M.~Lugaro, P.~Marigo, E.~Masha,
  R.~Menegazzo, V.~Mossa, F.~R. Pantaleo, V.~Paticchio, R.~Perrino, D.~Piatti,
  P.~Prati, L.~Schiavulli, K.~St{\"o}ckel, O.~Straniero, T.~Sz{\"u}cs, M.~P.
  Tak{\'a}cs, F.~Terrasi, D.~Trezzi, and S.~Zavatarelli.
\newblock A new approach to monitor ${}^{13}\mathrm{C}$-targets degradation in
  situ for ${}^{13}\mathrm{C}(\alpha, \mathrm{n}){}^{16}\mathrm{O}$
  cross-section measurements at luna.
\newblock {\em European Physical Journal A}, 56(3):75, 3 2020.

\bibitem{Boothroyd-95ApJ}
Arnold~I. {Boothroyd}, I.~Juliana {Sackmann}, and G.~J. {Wasserburg}.
\newblock {Hot Bottom Burning in Asymptotic Giant Branch Stars and Its Effect
  on Oxygen Isotopic Abundances}.
\newblock {\em Astrophys.\ J.\ Lett.}, 442:L21, 3 1995.

\bibitem{Herwig2005}
Falk {Herwig}.
\newblock {Evolution of Asymptotic Giant Branch Stars}.
\newblock {\em Annual Review of Astronomy and Astrophysics}, 43(1):435--479, 9
  2005.

\bibitem{Buchmann06}
L.~R. {Buchmann} and C.~A. {Barnes}.
\newblock {Nuclear reactions in stellar helium burning and later hydrostatic
  burning stages}.
\newblock {\em Nucl. Phys. A}, 777:254--290, 10 2006.

\bibitem{Kaeppeler2011}
F.~K\"appeler, R.~Gallino, S.~Bisterzo, and W.~Aoki.
\newblock The $s$ process: {N}uclear physics, stellar models, and observations.
\newblock {\em Rev.\ Mod.\ Phys.}, 83:157, 2011.

\bibitem{Carretta2009}
E.~{Carretta}, A.~{Bragaglia}, R.~G. {Gratton}, S.~{Lucatello}, G.~{Catanzaro},
  F.~{Leone}, M.~{Bellazzini}, R.~{Claudi}, V.~{D'Orazi}, Y.~{Momany},
  S.~{Ortolani}, E.~{Pancino}, G.~{Piotto}, A.~{Recio-Blanco}, and E.~{Sabbi}.
\newblock {Na-O anticorrelation and HB. VII. The chemical composition of first
  and second-generation stars in 15 globular clusters from GIRAFFE spectra}.
\newblock {\em Astronomy and Astrophysics}, 505(1):117--138, 10 2009.

\bibitem{Gratton2012}
Raffaele~G. {Gratton}, Eugenio {Carretta}, and Angela {Bragaglia}.
\newblock {Multiple populations in globular clusters. Lessons learned from the
  Milky Way globular clusters}.
\newblock {\em The Astronomy and Astrophysics Review}, 20:50, 2 2012.

\bibitem{Ventura2018}
P.~{Ventura}, A.~{Karakas}, F.~{Dell'Agli}, D.~A. {Garc{\'\i}a-Hern{\'a}ndez},
  and L.~{Guzman-Ramirez}.
\newblock {Gas and dust from solar metallicity AGB stars}.
\newblock {\em Monthly Notices of the Royal Astronomical Society},
  475(2):2282--2305, 4 2018.

\bibitem{Goerres82-NPA}
J.~{G{\"o}rres}, C.~{Rolfs}, P.~{Schmalbrock}, H.~P. {Trautvetter}, and
  J.~{Keinonen}.
\newblock {Search for low-energy resonances in
  ${}^{21}\mathrm{Ne}(\mathrm{p},\gamma){}^{22}\mathrm{Na}$ and
  ${}^{22}\mathrm{Ne}(\mathrm{p},\gamma){}^{23}\mathrm{Na}$}.
\newblock {\em Nucl. Phys. A}, 385(1):57--75, 8 1982.

\bibitem{ANGULO1999}
C.~Angulo, M.~Arnould, M.~Rayet, P.~Descouvemont, D.~Baye, C.~Leclercq-Willain,
  A.~Coc, S.~Barhoumi, P.~Aguer, C.~Rolfs, R.~Kunz, J.W. Hammer, A.~Mayer,
  T.~Paradellis, S.~Kossionides, C.~Chronidou, K.~Spyrou, S.~Degl'Innocenti,
  G.~Fiorentini, B.~Ricci, S.~Zavatarelli, C.~Providencia, H.~Wolters,
  J.~Soares, C.~Grama, J.~Rahighi, A.~Shotter, and M.~Lamehi Rachti.
\newblock A compilation of charged-particle induced thermonuclear reaction
  rates.
\newblock {\em Nucl. Phys. A}, 656(1):3 -- 183, 1999.

\bibitem{Sallaska13-ApJSS}
A.~L. {Sallaska}, C.~{Iliadis}, A.~E. {Champange}, S.~{Goriely},
  S.~{Starrfield}, and F.~X. {Timmes}.
\newblock {STARLIB: A Next-generation Reaction-rate Library for Nuclear
  Astrophysics}.
\newblock {\em Astrophys.\ J.\ Suppl.\ Ser.}, 207:18, 7 2013.

\bibitem{Cavanna2015}
F.~Cavanna, R.~Depalo, M.~Aliotta, M.~Anders, D.~Bemmerer, A.~Best,
  A.~Boeltzig, C.~Broggini, C.~G. Bruno, A.~Caciolli, P.~Corvisiero,
  T.~Davinson, A.~di~Leva, Z.~Elekes, F.~Ferraro, A.~Formicola, Zs. F\"ul\"op,
  G.~Gervino, A.~Guglielmetti, C.~Gustavino, Gy. Gy\"urky, G.~Imbriani,
  M.~Junker, R.~Menegazzo, V.~Mossa, F.~R. Pantaleo, P.~Prati, D.~A. Scott,
  E.~Somorjai, O.~Straniero, F.~Strieder, T.~Sz\"ucs, M.~P. Tak\'acs, and
  D.~Trezzi.
\newblock Three new low-energy resonances in the
  ${}^{22}\mathrm{Ne}(\mathrm{p},\gamma){}^{23}\mathrm{Na}$ reaction.
\newblock {\em Phys.\ Rev.\ Lett.}, 115:252501, 12 2015.

\bibitem{slemer2016}
A.~Slemer, P.~Marigo, D.~Piatti, M.~Aliotta, D.~Bemmerer, A.~Best, A.~Boeltzig,
  A.~Bressan, C.~Broggini, C.~G. Bruno, A.~Caciolli, F.~Cavanna, G.~F. Ciani,
  P.~Corvisiero, T.~Davinson, R.~Depalo, A.~Di~Leva, Z.~Elekes, F.~Ferraro,
  A.~Formicola, Zs. F\"ul\"op, G.~Gervino, A.~Guglielmetti, C.~Gustavino,
  G.~Gy\"urky, G.~Imbriani, M.~Junker, R.~Menegazzo, V.~Mossa, F.~R. Pantaleo,
  P.~Prati, O.~Straniero, T.~Sz\"ucs, M.~P. Tak\'acs, and D.~Trezzi.
\newblock {${}^{22}\mathrm{Ne}$ and ${}^{23}\mathrm{Na}$ ejecta from
  intermediate-mass stars: the impact of the new LUNA rate for
  ${}^{22}\mathrm{Ne}(\mathrm{p},\gamma){}^{23}\mathrm{Na}$}.
\newblock {\em Monthly Notices of the Royal Astronomical Society},
  465(4):4817--4837, 11 2016.

\bibitem{Iliadis10-NPA}
C.~{Iliadis}, R.~{Longland}, A.~E. {Champagne}, and A.~{Coc}.
\newblock {Charged-particle thermonuclear reaction rates: III. Nuclear physics
  input}.
\newblock {\em Nucl. Phys. A}, 841(1-4):251--322, 10 2010.

\bibitem{Kelly2017}
K.~J. Kelly, A.~E. Champagne, L.~N. Downen, J.~R. Dermigny, S.~Hunt,
  C.~Iliadis, and A.~L. Cooper.
\newblock New measurements of low-energy resonances in the
  $^{22}\mathbf{Ne}(p,\ensuremath{\gamma})^{23}\mathbf{Na}$ reaction.
\newblock {\em Phys.\ Rev.\ C}, 95:015806, Jan 2017.

\bibitem{Ferraro2018-PRL}
F.~Ferraro, M.~P. Tak\'acs, D.~Piatti, F.~Cavanna, R.~Depalo, M.~Aliotta,
  D.~Bemmerer, A.~Best, A.~Boeltzig, C.~Broggini, C.~G. Bruno, A.~Caciolli,
  T.~Chillery, G.~F. Ciani, P.~Corvisiero, T.~Davinson, G.~D'Erasmo,
  A.~Di~Leva, Z.~Elekes, E.~M. Fiore, A.~Formicola, Zs. F\"ul\"op, G.~Gervino,
  A.~Guglielmetti, C.~Gustavino, Gy. Gy\"urky, G.~Imbriani, M.~Junker,
  A.~Karakas, I.~Kochanek, M.~Lugaro, P.~Marigo, R.~Menegazzo, V.~Mossa, F.~R.
  Pantaleo, V.~Paticchio, R.~Perrino, P.~Prati, L.~Schiavulli, K.~St\"ockel,
  O.~Straniero, T.~Sz\"ucs, D.~Trezzi, and S.~Zavatarelli.
\newblock Direct capture cross section and the ${E}_\mathrm{p}=71$ and 105\,kev
  resonances in the ${}^{22}\mathrm{Ne}(\mathrm{p},\gamma)^{23}\mathrm{Na}$
  reaction.
\newblock {\em Phys.\ Rev.\ Lett.}, 121:172701, 10 2018.

\bibitem{Heil2008}
M.~Heil, R.~Detwiler, R.~E. Azuma, A.~Couture, J.~Daly, J.~G\"orres,
  F.~K\"appeler, R.~Reifarth, P.~Tischhauser, C.~Ugalde, and M.~Wiescher.
\newblock {The ${}^{13}\mathrm{C}(\alpha,\mathrm{n})$ reaction and its role as
  a neutron source for the $s$ process}.
\newblock {\em Phys.\ Rev.\ C}, 78:025803, Aug 2008.

\bibitem{Depalo15-PRC}
Rosanna Depalo, Francesca Cavanna, Federico Ferraro, Alessandra Slemer, Tariq
  Al-Abdullah, Shavkat Akhmadaliev, Michael Anders, Daniel Bemmerer, Zolt\'an
  Elekes, Giovanni Mattei, Stefan Reinicke, Konrad Schmidt, Carlo Scian, and
  Louis Wagner.
\newblock Strengths of the resonances at 436, 479, 639, 661, and 1279\,{keV} in
  the ${}^{22}\mathrm{Ne}(\mathrm{p},\gamma){}^{23}\mathrm{Na}$ reaction.
\newblock {\em Phys.\ Rev.\ C}, 92(4):045807, 10 2015.

\bibitem{Depalo2016}
R.~Depalo, F.~Cavanna, M.~Aliotta, M.~Anders, D.~Bemmerer, A.~Best,
  A.~Boeltzig, C.~Broggini, C.~G. Bruno, A.~Caciolli, G.~F. Ciani,
  P.~Corvisiero, T.~Davinson, A.~Di~Leva, Z.~Elekes, F.~Ferraro, A.~Formicola,
  Zs. F\"ul\"op, G.~Gervino, A.~Guglielmetti, C.~Gustavino, Gy. Gy\"urky,
  G.~Imbriani, M.~Junker, R.~Menegazzo, V.~Mossa, F.~R. Pantaleo, D.~Piatti,
  P.~Prati, O.~Straniero, T.~Sz\"ucs, M.~P. Tak\'acs, and D.~Trezzi.
\newblock Direct measurement of low-energy
  ${}^{22}\mathrm{Ne}(\mathrm{p},\gamma){}^{23}\mathrm{Na}$ resonances.
\newblock {\em Phys.\ Rev.\ C}, 94:055804, 11 2016.

\bibitem{Marion57-ApJ}
J.B. Marion and W.A. Fowler.
\newblock {Nuclear Reactions with the Neon Isotopes in Stars.}
\newblock {\em Astrophys.\ J.}, 125:221, 1 1957.

\bibitem{Gratton04-ARAA}
R.~{Gratton}, C.~{Sneden}, and E.~{Carretta}.
\newblock {Abundance Variations Within Globular Clusters}.
\newblock {\em Annual Review of Astronomy and Astrophysics}, 42:385--440, 9
  2004.

\bibitem{Renzini_2015}
A.~Renzini, F.~D’Antona, S.~Cassisi, I.~R. King, A.~P. Milone, P.~Ventura,
  J.~Anderson, L.~R. Bedin, A.~Bellini, T.~M. Brown, and et~al.
\newblock Thehubble space telescopeuv legacy survey of galactic globular
  clusters – v. constraints on formation scenarios.
\newblock {\em Monthly Notices of the Royal Astronomical Society},
  454(4):4197–4207, Oct 2015.

\bibitem{Ventura06-AA}
P.~{Ventura} and F.~{D'Antona}.
\newblock {Does the oxygen-sodium anticorrelation in globular clusters require
  a lowering of the $^{23}$Na(p,{$\alpha$})$^{20}$Ne reaction rate?}
\newblock {\em Astronomy and Astrophysics}, 457:995--1001, 10 2006.

\bibitem{Rolfs75-NPA}
C.~{Rolfs}, W.~S. {Rodney}, M.~H. {Shapiro}, and H.~{Winkler}.
\newblock {Hydrogen burning of ${}^{20}\mathrm{Ne}$ and ${}^{22}\mathrm{Ne}$ in
  stars}.
\newblock {\em Nucl. Phys. A}, 241:460--486, 4 1975.

\bibitem{Mukhamedzhanov}
A.~M. Mukhamedzhanov, P.~B\'em, V.~Burjan, C.~A. Gagliardi, B.~F. Irgaziev,
  V.~Kroha, J.~Nov\'ak, \v{S} Pisko\v{r}, E.~\v{S}ime\v{c}kov\'a, R.~E.
  Tribble, F.~Vesel\'y, and J.~Vincour.
\newblock Asymptotic normalization coefficients from the
  ${}^{20}\mathrm{Ne}({}^{3}\mathrm{He}, \mathrm{d}){}^{21}\mathrm{Na}$
  reaction and astrophysical factor for
  ${}^{20}\mathrm{Ne}(\mathrm{p},\gamma){}^{21}\mathrm{Na}$.
\newblock {\em Phys.\ Rev.\ C}, 73:035806, Mar 2006.

\bibitem{Lyons2018}
S.~Lyons, J.~G\"orres, R.~J. deBoer, E.~Stech, Y.~Chen, G.~Gilardy, Q.~Liu,
  A.~M. Long, M.~Moran, D.~Robertson, C.~Seymour, B.~Vande~Kolk, M.~Wiescher,
  and A.~Best.
\newblock Determination of
  ${}^{20}\mathrm{Ne}(\mathrm{p},\gamma)^{21}\mathrm{Na}$ cross sections from
  ${E}_\mathrm{p} = 500 - 2000\,\mathrm{keV}$.
\newblock {\em Phys.\ Rev.\ C}, 97:065802, Jun 2018.

\bibitem{Bemmerer18-EPL}
D.~Bemmerer, F.~Cavanna, R.~Depalo, M.~Aliotta, M.~Anders, A.~Boeltzig,
  C.~Broggini, C.~Bruno, A.~Caciolli, T.~Chillery, P.~Corvisiero, T.~Davinson,
  Z.~Elekes, F.~Ferraro, A.~Formicola, Zs. F\"ul\"op, G.~Gervino,
  A.~Guglielmetti, C.~Gustavino, Gy. Gy\"urky, R.~Menegazzo, V.~Mossa, F.~R.
  Pantaleo, P.~Prati, D.~A. Scott, K.~St\"ocker, O.~Straniero, T.~Sz\"ucs,
  M.~P. Tak\'acs, and D.~Trezzi.
\newblock Effect of beam energy straggling on resonant yield in thin gas
  targets: The cases ${}^{22}\mathrm{Ne}(\mathrm{p},\gamma){}^{23}\mathrm{Na}$
  and ${}^{14}\mathrm{N}(\mathrm{p},\gamma){}^{15}\mathrm{O}$.
\newblock {\em Europhysics Letters}, 122(5):52001, 2018.

\bibitem{Goerres1989}
J.~G\"orres, M.~Wiescher, and C.~Rolfs.
\newblock Hydrogen burning of ${}^{23}\mathrm{Na}$ in the {NeNa} cycle.
\newblock {\em Astrophys.\ J.}, 343:365, 8 1989.

\bibitem{Rowland2004}
C.~Rowland, C.~Iliadis, A.~E. Champagne, C.~Fox, J.~José, and R.~Runkle.
\newblock Does an {NeNa} {Cycle} {Exist} in {Explosive} {Hydrogen} {Burning}?
\newblock {\em Astrophys.\ J.}, 615(1):L37--L40, 11 2004.

\bibitem{Cesaratto2013}
J.~M. Cesaratto, A.~E. Champagne, M.~Q. Buckner, T.~B. Clegg, S.~Daigle,
  C.~Howard, C.~Iliadis, R.~Longland, J.~R. Newton, and B.~M. Oginni.
\newblock Measurement of the $e_\mathrm{r}^\mathrm{c.m.} = 138\,\mathrm{keV}$
  resonance in the ${}^{23}\mathrm{Na}(\mathrm{p},\gamma){}^{24}\mathrm{Mg}$
  reaction and the abundance of sodium in {AGB} stars.
\newblock {\em Phys.\ Rev.\ C}, 88(6):065806, 12 2013.

\bibitem{Boeltzig2019}
A.~Boeltzig, A.~Best, F.R. Pantaleo, G.~Imbriani, M.~Junker, M.~Aliotta,
  J.~Balibrea-Correa, D.~Bemmerer, C.~Broggini, C.G. Bruno, R.~Buompane,
  A.~Caciolli, F.~Cavanna, T.~Chillery, G.F. Ciani, P.~Corvisiero, L.~Csedreki,
  T.~Davinson, R.J. deBoer, R.~Depalo, A.~Di~Leva, Z.~Elekes, F.~Ferraro, E.M.
  Fiore, A.~Formicola, Zs. F\"ul\"op, G.~Gervino, A.~Guglielmetti,
  C.~Gustavino, Gy. Gy\"urky, I.~Kochanek, M.~Lugaro, P.~Marigo, R.~Menegazzo,
  V.~Mossa, F.~Munnik, V.~Paticchio, R.~Perrino, D.~Piatti, P.~Prati,
  L.~Schiavulli, K.~St\"ockel, O.~Straniero, F.~Strieder, T.~Sz\"ucs, M.P.
  Tak\'acs, D.~Trezzi, M.~Wiescher, and S.~Zavatarelli.
\newblock Direct measurements of low-energy resonance strengths of the
  ${}^{23}\mathrm{Na}(\mathrm{p},\gamma){}^{24}\mathrm{Mg}$ reaction for
  astrophysics.
\newblock {\em Phys.\ Lett.\ B}, 795:122--128, 8 2019.

\bibitem{Marshall2021}
C.~Marshall, K.~Setoodehnia, F.~Portillo, J.~H. Kelley, and R.~Longland.
\newblock New energy for the 133-{keV} resonance in the
  ${}^{23}\mathrm{Na}(\mathrm{p},\gamma){}^{24}\mathrm{Mg}$ reaction and its
  impact on nucleosynthesis in globular clusters.
\newblock {\em Phys.\ Rev.\ C}, 104(3):L032801, 9 2021.

\bibitem{Hale2004}
S.E. Hale, A.~Champagne, C.~Iliadis, V.~Hansper, D.~Powell, and J.~Blackmon.
\newblock Investigation of the
  ${}^{23}\mathrm{Na}(\mathrm{p},\gamma){}^{24}\mathrm{Mg}$ and
  ${}^{23}\mathrm{Na}(\mathrm{p},\alpha){}^{20}\mathrm{Ne}$ reactions via
  $({}^{3}\mathrm{He},\mathrm{d})$ spectroscopy.
\newblock {\em Phys.\ Rev.\ C}, 70(4):045802, 10 2004.

\bibitem{Straniero2006}
Oscar Straniero, Roberto Gallino, and Sergio Cristallo.
\newblock s process in low-mass asymptotic giant branch stars.
\newblock {\em Nucl. Phys. A}, 777:311, 2006.

\bibitem{Pignatari2010}
M.~Pignatari, R.~Gallino, M.~Heil, M.~Wiescher, F.~Käppeler, F.~Herwig, and
  S.~Bisterzo.
\newblock {THE} {WEAKs}-{PROCESS} {IN} {MASSIVE} {STARS} {AND} {ITS}
  {DEPENDENCE} {ON} {THE} {NEUTRON} {CAPTURE} {CROSS} {SECTIONS}.
\newblock {\em Astrophys.\ J.}, 710(2):1557--1577, feb 2010.

\bibitem{Busso01-APJ}
Maurizio {Busso}, Roberto {Gallino}, David~L. {Lambert}, Claudia {Travaglio},
  and Verne~V. {Smith}.
\newblock {Nucleosynthesis and Mixing on the Asymptotic Giant Branch. III.
  Predicted and Observed s-Process Abundances}.
\newblock {\em Astrophys.\ J.}, 557(2):802--821, 8 2001.

\bibitem{Busso2021}
Maurizio Busso, Diego Vescovi, Sara Palmerini, Sergio Cristallo, and Vincenzo
  Antonuccio-Delogu.
\newblock s-processing in {AGB} stars revisited. {III}. neutron captures from
  {MHD} mixing at different metallicities and observational constraints.
\newblock {\em Astrophys.\ J.}, 908(1):55, feb 2021.

\bibitem{gallino1998}
Roberto {Gallino}, Claudio {Arlandini}, Maurizio {Busso}, Maria {Lugaro},
  Claudia {Travaglio}, Oscar {Straniero}, Alessandro {Chieffi}, and Marco
  {Limongi}.
\newblock {Evolution and Nucleosynthesis in Low-Mass Asymptotic Giant Branch
  Stars. II. Neutron Capture and the $s$-Process}.
\newblock {\em Astrophys.\ J.}, 497(1):388--403, 4 1998.

\bibitem{Brune1993}
C.~R. Brune, I.~Licot, and R.~W. Kavanagh.
\newblock Low-energy resonances in ${}^{13}\mathrm{C}(\alpha,\mathrm{n})$.
\newblock {\em Phys.\ Rev.\ C}, 48:3119--3121, Dec 1993.

\bibitem{Drotleff1993}
H.~W. {Drotleff}, A.~{Denker}, H.~{Knee}, M.~{Soine}, G.~{Wolf}, J.~W.
  {Hammer}, U.~{Greife}, C.~{Rolfs}, and H.~P. {Trautvetter}.
\newblock {{Reaction rates of the s-process neutron sources
  ${}^{22}\mathrm{Ne}(\alpha,\mathrm{n}){}^{25}\mathrm{Mg}$ and
  ${}^{13}\mathrm{C}(\alpha,\mathrm{n}){}^{16}\mathrm{O}$}}.
\newblock {\em Astrophys.\ J.}, 414:735, sep 1993.

\bibitem{Guo2012}
B.~{Guo}, Z.~H. {Li}, M.~{Lugaro}, J.~{Buntain}, D.~Y. {Pang}, Y.~J. {Li},
  J.~{Su}, S.~Q. {Yan}, X.~X. {Bai}, Y.~S. {Chen}, Q.~W. {Fan}, S.~J. {Jin},
  A.~I. {Karakas}, E.~T. {Li}, Z.~C. {Li}, G.~{Lian}, J.~C. {Liu}, X.~{Liu},
  J.~R. {Shi}, N.~C. {Shu}, B.~X. {Wang}, Y.~B. {Wang}, S.~{Zeng}, and W.~P.
  {Liu}.
\newblock {New Determination of the
  ${}^{13}\mathrm{C}(\alpha,\mathrm{n})^{16}\mathrm{O}$ Reaction Rate and its
  Influence on the $s$-process Nucleosynthesis in AGB Stars}.
\newblock {\em Astrophys.\ J.}, 756(2):193, 9 2012.

\bibitem{Avila2015}
M.~L. Avila, G.~V. Rogachev, E.~Koshchiy, L.~T. Baby, J.~Belarge, K.~W. Kemper,
  A.~N. Kuchera, and D.~Santiago-Gonzalez.
\newblock {New measurement of the $\alpha$ asymptotic normalization coefficient
  of the ${1/2}^{+}$ state in ${}^{17}\mathrm{O}$ at 6.356\,MeV that dominates
  the ${}^{13}\mathrm{C}(\alpha,\mathrm{n})^{16}\mathrm{O}$ reaction rate at
  temperatures relevant for the $s$ process}.
\newblock {\em Phys.\ Rev.\ C}, 91:048801, Apr 2015.

\bibitem{Trippella2017}
O.~Trippella and M.~La Cognata.
\newblock {Concurrent Application of ANC and THM to assess the
  ${}^{13}\mathrm{C}(\alpha,\mathrm{n})^{16}\mathrm{O}$ Absolute Cross Section
  at Astrophysical Energies and Possible Consequences for Neutron Production in
  Low-mass AGB Stars}.
\newblock {\em Astrophys.\ J.}, 837(1):41, 2017.

\bibitem{Cristallo2018}
S.~Cristallo, M.~La Cognata, C.~Massimi, A.~Best, S.~Palmerini, O.~Straniero,
  O.~Trippella, M.~Busso, G.~F. Ciani, F.~Mingrone, L.~Piersanti, and
  D.~Vescovi.
\newblock {The Importance of the
  ${}^{13}\mathrm{C}(\alpha,\mathrm{n})^{16}\mathrm{O}$ Reaction in Asymptotic
  Giant Branch Stars}.
\newblock {\em Astrophys.\ J.}, 859(2):105, 2018.

\bibitem{DeBoer2020}
R.~J. deBoer, C.~R. Brune, M.~Febrarro, J.~G\"orres, I.~J. Thompson, and
  M.~Wiescher.
\newblock Sensitivity of the
  ${}^{13}\mathrm{C}(\alpha,\mathrm{n})^{16}\mathrm{O}$ $s$ factor to the
  uncertainty in the level parameters of the near-threshold state.
\newblock {\em Phys.\ Rev.\ C}, 101:045802, Apr 2020.

\bibitem{Ciani2021}
G.~F. Ciani, L.~Csedreki, D.~Rapagnani, M.~Aliotta, J.~Balibrea-Correa,
  F.~Barile, D.~Bemmerer, A.~Best, A.~Boeltzig, C.~Broggini, C.~G. Bruno,
  A.~Caciolli, F.~Cavanna, T.~Chillery, P.~Colombetti, P.~Corvisiero,
  S.~Cristallo, T.~Davinson, R.~Depalo, A.~Di~Leva, Z.~Elekes, F.~Ferraro,
  E.~Fiore, A.~Formicola, Zs. F\"ul\"op, G.~Gervino, A.~Guglielmetti,
  C.~Gustavino, Gy. Gy\"urky, G.~Imbriani, M.~Junker, M.~Lugaro, P.~Marigo,
  E.~Masha, R.~Menegazzo, V.~Mossa, F.~R. Pantaleo, V.~Paticchio, R.~Perrino,
  D.~Piatti, P.~Prati, L.~Schiavulli, K.~St\"ockel, O.~Straniero, T.~Sz\"ucs,
  M.~P. Tak\'acs, F.~Terrasi, D.~Vescovi, and S.~Zavatarelli.
\newblock Direct measurement of the
  ${}^{13}\mathrm{C}(\alpha,\mathrm{n}){}^{16}\mathrm{O}$ cross section into
  the $s$-process gamow peak.
\newblock {\em Phys.\ Rev.\ Lett.}, 127:152701, Oct 2021.

\bibitem{Azuma2010}
R.E. Azuma, E.~Uberseder, E.C. Simpson, C.R. Brune, H.~Costantini, R.J.
  de~Boer, J.~G\"orres, M.~Heil, P.J. LeBlanc, C.~Ugalde, and M.~Wiescher.
\newblock {AZURE}: {A}n $r$-matrix code for nuclear astrophysics.
\newblock {\em Phys.\ Rev.\ C}, 81:045805, 2010.

\bibitem{Harissopulos2005}
S.~Harissopulos, H.~W. Becker, J.~W. Hammer, A.~Lagoyannis, C.~Rolfs, and
  F.~Strieder.
\newblock Cross section of the
  ${}^{13}\mathrm{C}(\alpha,\mathrm{n}){}^{16}\mathrm{O}$ reaction: {A}
  background for the measurement of geo-neutrinos.
\newblock {\em Phys.\ Rev.\ C}, 72:062801(R), 2005.

\bibitem{Xu2013}
Y.~{Xu}, K.~{Takahashi}, S.~{Goriely}, M.~{Arnould}, M.~{Ohta}, and
  H.~{Utsunomiya}.
\newblock {NACRE II: an update of the NACRE compilation of
  charged-particle-induced thermonuclear reaction rates for nuclei with mass
  number $A < 16$}.
\newblock {\em Nucl. Phys. A}, 918:61--169, 11 2013.

\bibitem{Thielemann-2018ASSL}
Friedrich-Karl {Thielemann}, Roland {Diehl}, Alexander {Heger}, Raphael
  {Hirschi}, and Matthias {Liebendoerfer}.
\newblock {Massive Stars and Their Supernovae}.
\newblock In Roland {Diehl}, Dieter~H. {Hartmann}, and Nikos {Prantzos},
  editors, {\em Astrophysics with Radioactive Isotopes}, volume 453 of {\em
  Astrophysics and Space Science Library}, pages 173--286. Springer, Cham,
  2018.

\bibitem{Lugaro-2018ASSL}
Maria {Lugaro} and Alessandro {Chieffi}.
\newblock {Low- and Intermediate-Mass Stars}.
\newblock In Roland {Diehl}, Dieter~H. {Hartmann}, and Nikos {Prantzos},
  editors, {\em Astrophysics with Radioactive Isotopes}, volume 453 of {\em
  Astrophysics and Space Science Library}, pages 91--172. Springer, Cham, 2018.

\bibitem{Karakas06-ApJ}
A.~I. {Karakas}, M.~A. {Lugaro}, M.~{Wiescher}, J.~{G{\"o}rres}, and
  C.~{Ugalde}.
\newblock {The Uncertainties in the ${}^{22}\mathrm{Ne}+\alpha$-Capture
  Reaction Rates and the Production of the Heavy Magnesium Isotopes in
  Asymptotic Giant Branch Stars of Intermediate Mass}.
\newblock {\em Astrophys.\ J.}, 643:471--483, 5 2006.

\bibitem{Adsley2021}
Philip Adsley, Umberto Battino, Andreas Best, Antonio Caciolli, Alessandra
  Guglielmetti, Gianluca Imbriani, Heshani Jayatissa, Marco La~Cognata, Livio
  Lamia, Eliana Masha, Cristian Massimi, Sara Palmerini, Ashley Tattersall, and
  Raphael Hirschi.
\newblock Reevaluation of the
  ${}^{22}\mathrm{Ne}(\alpha,\gamma){}^{26}\mathrm{Mg}$ and
  ${}^{22}\mathrm{Ne}(\alpha,\mathrm{n}){}^{25}\mathrm{Mg}$ reaction rates.
\newblock {\em Phys.\ Rev.\ C}, 103:015805, Jan 2021.

\bibitem{Jaeger01-PRL}
M.~{Jaeger}, R.~{Kunz}, A.~{Mayer}, J.~W. {Hammer}, G.~{Staudt}, K.~L. {Kratz},
  and B.~{Pfeiffer}.
\newblock {${}^{22}\mathrm{Ne}(\alpha,\mathrm{n}){}^{25}\mathrm{Mg}$: The Key
  Neutron Source in Massive Stars}.
\newblock {\em Phys.\ Rev.\ Lett.}, 87:202501, 2001.

\bibitem{Giesen93-NPA}
U.~{Giesen}, C.~P. {Browne}, J.~{G{\"o}rres}, S.~{Graff}, C.~{Iliadis}, H.-P.
  {Trautvetter}, M.~{Wiescher}, W.~{Harms}, K.~L. {Kratz}, B.~{Pfeiffer}, R.~E.
  {Azuma}, M.~{Buckby}, and J.~D. {King}.
\newblock {The astrophysical implications of low-energy resonances in
  ${}^{22}\mathrm{Ne} + \alpha$}.
\newblock {\em Nucl. Phys. A}, 561:95--111, 8 1993.

\bibitem{Ugalde07-PRC}
C.~{Ugalde}, A.~E. {Champagne}, S.~{Daigle}, C.~{Iliadis}, R.~{Longland}, J.~R.
  {Newton}, E.~{Osenbaugh-Stewart}, J.~A. {Clark}, C.~{Deibel}, A.~{Parikh},
  P.~D. {Parker}, and C.~{Wrede}.
\newblock {Experimental evidence for a natural parity state in
  ${}^{26}\mathrm{Mg}$ and its impact on the production of neutrons for the $s$
  process}.
\newblock {\em Phys.\ Rev.\ C}, 76(2):025802, 8 2007.

\bibitem{Longland09-PRC}
R.~{Longland}, C.~{Iliadis}, G.~{Rusev}, A.~P. {Tonchev}, R.~J. {Deboer},
  J.~{G{\"o}rres}, and M.~{Wiescher}.
\newblock {Photoexcitation of astrophysically important states in
  ${}^{26}\mathrm{Mg}$}.
\newblock {\em Phys.\ Rev.\ C}, 80(5):055803, 11 2009.

\bibitem{Talwar16-PRC}
R.~{Talwar}, T.~{Adachi}, G.~P.~A. {Berg}, L.~{Bin}, S.~{Bisterzo},
  M.~{Couder}, R.~J. {deBoer}, X.~{Fang}, H.~{Fujita}, Y.~{Fujita},
  J.~{G{\"o}rres}, K.~{Hatanaka}, T.~{Itoh}, T.~{Kadoya}, A.~{Long}, K.~{Miki},
  D.~{Patel}, M.~{Pignatari}, Y.~{Shimbara}, A.~{Tamii}, M.~{Wiescher},
  T.~{Yamamoto}, and M.~{Yosoi}.
\newblock Probing astrophysically important states in the ${}^{26}\mathrm{Mg}$
  nucleus to study neutron sources for the $s$ process.
\newblock {\em Phys.\ Rev.\ C}, 93:055803, May 2016.

\bibitem{Lotay19-EPJA}
G.~{Lotay}, D.~T. {Doherty}, D.~{Seweryniak}, S.~{Almaraz-Calderon}, M.~P.
  {Carpenter}, C.~J. {Chiara}, H.~M. {David}, C.~R. {Hoffman}, R.~V.~F.
  {Janssens}, A.~{Kankainen}, T.~{Lauritsen}, R.~{Wilkinson}, P.~J. {Woods},
  and S.~{Zhu}.
\newblock {Identification of $\gamma$-decaying resonant states in
  ${}^{26}\mathrm{Mg}$ and their importance for the astrophysical $s$ process}.
\newblock {\em European Physical Journal A}, 55(7):109, 7 2019.

\bibitem{Jayatissa-2020PhLB}
H.~{Jayatissa}, G.~V. {Rogachev}, V.~Z. {Goldberg}, E.~{Koshchiy},
  G.~{Christian}, J.~{Hooker}, S.~{Ota}, B.~T. {Roeder}, A.~{Saastamoinen},
  O.~{Trippella}, S.~{Upadhyayula}, and E.~{Uberseder}.
\newblock {Constraining the
  ${}^{22}\mathrm{Ne}(\alpha,\gamma){}^{26}\mathrm{Mg}$ and
  ${}^{22}\mathrm{Ne}(\alpha,\mathrm{n}){}^{25}\mathrm{Mg}$ reaction rates
  using sub-Coulomb $\alpha$-transfer reactions}.
\newblock {\em Phys.\ Lett.\ B}, 802:135267, 3 2020.

\bibitem{Ota-2020PhLB}
S.~Ota, G.~Christian, G.~Lotay, W.N. Catford, E.A. Bennett, S.~Dede, D.T.
  Doherty, S.~Hallam, J.~Hooker, C.~Hunt, H.~Jayatissa, A.~Matta, M.~Moukaddam,
  G.V. Rogachev, A.~Saastamoinen, J.A. Tostevin, S.~Upadhyayula, and
  R.~Wilkinson.
\newblock {Decay properties of ${}^{22}\mathrm{Ne} + \alpha$ resonances and
  their impact on $s$-process nucleosynthesis}.
\newblock {\em Phys.\ Lett.\ B}, 802:135256, 3 2020.

\bibitem{Longland12-PRC}
R.~{Longland}, C.~{Iliadis}, and A.~I. {Karakas}.
\newblock {Reaction rates for the $s$-process neutron source
  ${}^{22}\mathrm{Ne} + \alpha$}.
\newblock {\em Phys.\ Rev.\ C}, 85(6):065809, 6 2012.

\bibitem{Piatti18-PhD}
Denise Piatti.
\newblock {\em The Study of
  ${}^{22}\mathrm{Ne}(\alpha,\gamma){}^{26}\mathrm{Mg}$ and
  ${}^6\mathrm{Li}(\mathrm{p},\gamma){}^7\mathrm{Be}$ Reactions at LUNA}.
\newblock PhD thesis, Universit\`a degli studi di Padova, December 2018.

\bibitem{Bemmerer-2005EPJA}
D.~{Bemmerer}, F.~{Confortola}, A.~{Lemut}, R.~{Bonetti}, C.~{Broggini},
  P.~{Corvisiero}, H.~{Costantini}, J.~{Cruz}, A.~{Formicola}, Zs.
  {F{\"u}l{\"o}p}, G.~{Gervino}, A.~{Guglielmetti}, C.~{Gustavino}, Gy.
  {Gy{\"u}rky}, G.~{Imbriani}, A.~P. {Jesus}, M.~{Junker}, B.~{Limata},
  R.~{Menegazzo}, P.~{Prati}, V.~{Roca}, D.~{Rogalla}, C.~{Rolfs}, M.~{Romano},
  C.~{Rossi Alvarez}, F.~{Sch{\"u}mann}, E.~{Somorjai}, O.~{Straniero},
  F.~{Strieder}, F.~{Terrasi}, H.~P. {Trautvetter}, and A.~{Vomiero}.
\newblock {Feasibility of low-energy radiative-capture experiments at the LUNA
  underground accelerator facility}.
\newblock {\em European Physical Journal A}, 24(2):313--319, 5 2005.

\bibitem{Best16-EPJA}
{Best, A.}, {Caciolli, A.}, {F\"ul\"up, Zs.}, {Gy\"orky, Gy.}, {Laubenstein,
  M.}, {Napolitani, E.}, {Rigato, V.}, {Roca, V.}, and {Sz\"ucs, T.}
\newblock Underground nuclear astrophysics: {W}hy and how.
\newblock {\em European Physical Journal A}, 52(4):72, 2016.

\bibitem{Harms91}
V.~{Harms}, K.~L {Kratz}, and M.~{Wiescher}.
\newblock {Properties of
  ${}^{22}\mathrm{Ne}(\alpha,\mathrm{n}){}^{25}\mathrm{Mg}$ resonances }.
\newblock {\em Phys.\ Rev.\ C}, 43(6):2849, 1991.

\bibitem{Bisterzo_2014}
S.~Bisterzo, C.~Travaglio, R.~Gallino, M.~Wiescher, and F.~K\"appeler.
\newblock Galactic chemical evolution and solar $s$-process abundances:
  Dependence on the ${}^{13}\mathrm{C}$-pocket structure.
\newblock {\em Astrophys.\ J.}, 787(1):10, Apr 2014.

\bibitem{massimi12}
C.~Massimi, P.~Koehler, S.~Bisterzo, N.~Colonna, R.~Gallino, F.~Gunsing,
  F.~K\"appeler, G.~Lorusso, A.~Mengoni, M.~Pignatari, G.~Vannini,
  U.~Abbondanno, G.~Aerts, H.~\'Alvarez, F.~\'Alvarez-Velarde, S.~Andriamonje,
  J.~Andrzejewski, P.~Assimakopoulos, L.~Audouin, G.~Badurek, M.~Barbagallo,
  P.~Baumann, F.~Be\ifmmode~\check{c}\else \v{c}\fi{}v\'a\v{r}, F.~Belloni,
  M.~Bennett, E.~Berthoumieux, M.~Calviani, F.~Calvi\~no, D.~Cano-Ott,
  R.~Capote, C.~Carrapi\ifmmode~\mbox{\c{c}}\else \c{c}\fi{}o, A.~Carrillo~de
  Albornoz, P.~Cennini, V.~Chepel, E.~Chiaveri, G.~Cortes, A.~Couture, J.~Cox,
  M.~Dahlfors, S.~David, I.~Dillmann, R.~Dolfini, C.~Domingo-Pardo, W.~Dridi,
  I.~Duran, C.~Eleftheriadis, M.~Embid-Segura, L.~Ferrant, A.~Ferrari,
  R.~Ferreira-Marques, L.~Fitzpatrick, H.~Frais-Koelbl, K.~Fujii, W.~Furman,
  I.~Goncalves, E.~Gonz\'alez-Romero, A.~Goverdovski, F.~Gramegna,
  E.~Griesmayer, C.~Guerrero, B.~Haas, R.~Haight, M.~Heil, A.~Herrera-Martinez,
  F.~Herwig, R.~Hirschi, M.~Igashira, S.~Isaev, E.~Jericha, Y.~Kadi,
  D.~Karadimos, D.~Karamanis, M.~Kerveno, V.~Ketlerov, V.~Konovalov,
  S.~Kopecky, E.~Kossionides, M.~Krti\ifmmode~\check{c}\else \v{c}\fi{}ka,
  C.~Lampoudis, H.~Leeb, C.~Lederer, A.~Lindote, I.~Lopes, R.~Losito,
  M.~Lozano, S.~Lukic, J.~Marganiec, L.~Marques, S.~Marrone, T.~Mart\'{\i}nez,
  P.~Mastinu, E.~Mendoza, P.~M. Milazzo, C.~Moreau, M.~Mosconi, F.~Neves,
  H.~Oberhummer, S.~O'Brien, M.~Oshima, J.~Pancin, C.~Papachristodoulou,
  C.~Papadopoulos, C.~Paradela, N.~Patronis, A.~Pavlik, P.~Pavlopoulos,
  L.~Perrot, M.~T. Pigni, R.~Plag, A.~Plompen, A.~Plukis, A.~Poch, J.~Praena,
  C.~Pretel, J.~Quesada, T.~Rauscher, R.~Reifarth, G.~Rockefeller, M.~Rosetti,
  C.~Rubbia, G.~Rudolf, J.~Salgado, C.~Santos, L.~Sarchiapone, R.~Sarmento,
  I.~Savvidis, C.~Stephan, G.~Tagliente, J.~L. Tain, D.~Tarr\'{\i}o,
  L.~Tassan-Got, L.~Tavora, R.~Terlizzi, P.~Vaz, A.~Ventura, D.~Villamarin,
  V.~Vlachoudis, R.~Vlastou, F.~Voss, S.~Walter, H.~Wendler, M.~Wiescher, and
  K.~Wisshak.
\newblock Resonance neutron-capture cross sections of stable magnesium isotopes
  and their astrophysical implications.
\newblock {\em Phys.\ Rev.\ C}, 85:044615, Apr 2012.

\end{thebibliography}

\end{document}